\documentclass[%
prl, twocolumn,
superscriptaddress,
10pt, 
 amsmath,amssymb,aps,
]{revtex4-2}

\usepackage{graphicx}
\usepackage{dcolumn}
\usepackage{bm}
\usepackage{xcolor}
\usepackage{color,soul} 
\usepackage{physics}
\usepackage[normalem]{ulem}
\usepackage{ytableau} 

\usepackage{amsmath}
\usepackage{amssymb}
\usepackage{natbib}

\usepackage[percent]{overpic}
\usepackage{dcolumn}%

\usepackage[dvipsnames,table,xcdraw]{xcolor}

\usepackage{bm}
\usepackage{braket}
\usepackage{dsfont}
\newcommand{\Lower}[1]{\underline{#1}}
\newcommand{\Upper}[1]{\overline{#1}}

\begin{document}

\preprint{APS/123-QED}

\title{A Generalized Coherent State Framework for Many-Body Density of States
}

\author{Deniz Coskun}
\affiliation{Institute for Theoretical Physics, ETHZ.
}%
\author{R. Chitra}%
\affiliation{Institute for Theoretical Physics, ETHZ.
}%




\date{\today}

\begin{abstract}

We develop a general framework to calculate the many-body density of states (DOS) of isolated and interacting quantum systems. Based on the  generalized coherent state formalism and the Simon-Lieb bounds for a quantum partition function, our method  provides a general method of calculation for the DOS in high-dimensional irreducible sectors. This framework further provides rigorous bounds for the ground state energy in each sector and enables the calculation of  microcanonical observables across the entire spectrum. Using the Lipkin-Meshkov-Glick (LMG) model  as a test bed, we validate our framework  by successfully identifying quantum phase transitions (QPTs) and excited-state quantum phase transitions (ESQPTs) across  spin sectors.   We then study  the ferromagnetic transverse field Ising chain with power law interactions,  where the hamiltonian ceases to be block diagonal in irreducible sectors. We generalize our coherent states to explicitly track the permutation symmetry group. For the strongly long range interacting case, we  demonstrate that the DOS in highest-spin sectors is qualitatively identical to that of  LMG-type Hamiltonians in the thermodynamic limit.
Unlike existing model-specific numerical or analytical techniques, our formalism relies on general underlying symmetries, making it broadly applicable.  It  establishes a versatile and computationally efficient bridge between algebraic structure and many-body thermodynamics.
\end{abstract}

\maketitle
\setcounter{secnumdepth}{3}
\section{\label{sec:Intro} Introduction}
The  study of     ground states and   low energy excitations  lies at the core of conventional quantum many-body  physics. 
However, the advent of  diverse quantum simulator platforms  such as trapped ions, Rydberg atom arrays, and superconducting circuits \cite{FossFeig2021TrappedIonSim, Browaeys2020ManyBodyRydberg, Kjaergaard2020SuperconductingQubits, ScienceUltraCold_2017, Arute2019QuantumSupremacy} permit the controlled exploration of  physics out of equilibrium, which  encode aspects of  the full energy spectrum.   
The exploration of the full manybody spectrum unveiled novel   phenomena  like   many-body localization\cite{Nandkishore2015MBLReview, AbaninThermalization2019},  the emergence of quantum scars   and  excited-State Quantum Phase Transition (ESQPT) \cite{ManyBodyScarsReview2025, Cejnar_2021}. Unlike standard ground-state transitions, an ESQPT manifests as non-analytical singularities  in the manybody density of states (DOS)and its derivatives. ESQPTs  dictate the "speed limit" of quantum evolution and signal fundamental structural changes in the system’s symmetry\cite{RelaxSpeedPerez_2011}. ESQPTs play an important role in dynamical phase transitions\cite{Heyl_2018, Van_Damme_2023}  and in the context of  the Eigenstate Thermalization Hypothesis (ETH), they can signal a break of  thermalization \cite{RelaxSpeedPerez_2011}.

Despite their significance, calculating the DOS for interacting many-body models remains a formidable challenge. Standard numerical techniques, such as exact diagonalization, are encumbered by the exponential growth of the Hilbert space \cite{SciPostPhysCodeb.70}, while analytical approaches, when possible, are largely confined to highly specific, integrable models \cite{Urbina_2023}.  The latter are based on  semiclassical methods which rely on the correspondence between   quantum operators and  scalar functions on a  phase space. Such methods have been  used   in the study of excited state and dynamical phase transitions \cite{Cejnar_2021, Mazza_2012, Sciolla_2013},  quantum chaos and many body scars \cite{ManyBodyScarsReview2025, Haake2018} and in the calculations of DOS in  models  with  all-to-all  interactions like  Lipkin-Meshkov-Glick (LMG) models and Dicke models,  \cite{Ribeiro_2008, Castanos2006, corps2024, Brandes2013}. The inherent complexity   of these approaches limits the DOS calculations to  highest quantum number sectors  and makes extensions to other models and quantum number sectors intractable.


In this work, we address this gap by presenting a general framework for calculating the DOS of an isolated interacting quantum system with a symmetry group $G $. Our method formalizes the heuristic approaches  of \cite{Mazza_2012, Sciolla_2013, Magnani_2014},  
and  is rooted in the rigorous bridge between quantum and classical mechanics established by Lieb \cite{Lieb1973} and  Simon \cite{Simon1980} which relates the Lie Group representations of quantum mechanics to the corresponding  symplectic geometry of classical mechanics, quantum operators onto scalar functions on a symplectic manifold. Using the generalized coherent states formalism\cite{Perelomov1986}, Lieb\cite{Lieb1973} and  Simon\cite{Simon1980} obtained  semiclassical bounds for the quantum canonical partition function of a  quantum system.   We demonstrate that these bounds can be harnessed to obtain  two estimates on the density of states (DOS) of an isolated quantum system in different quantum number sectors. These estimates converge with increasing quantum number, yielding the well defined DOS of the system. Using a generalized Hellmann-Feynman scheme, we show that  the bounds on the DOS can be  leveraged to derive microcanonical expectation values. This provides a powerful tool for calculating observables across the entire spectrum and establishing exact bounds on ground-state energies.

The paper is structured as follows: In Section \ref{sec:General Method}, we introduce our general method of DOS calculation. In Section \ref{sec:LMG-Model}, we provide a proof of principle by applying  our formalism   to the Lipkin-Meshkov-Glick (LMG) model, reproducing known analytical and numerical results. Finally, in Section \ref{sec: Ising}, we apply our method to the long-range interacting 1D Ising chain in a  transverse field  whose   Hamiltonian is no longer block diagonal in the individual irreps. Via  the use of additional coherent state bases, we demonstrate the versatility of our approach  beyond the standard LMG like models with all to all couplings.

\section{\label{sec:General Method}DOS Calculation Using Generalized Coherent States}
We present our method  based on generalized coherent states to calculate the DOS of a quantum system characterized by an irreducible high dimensional representation of a group $G$. By this we mean that the Hamiltonian acts on the representation, but not necessarily commutes with it.   Such a  coherent state based construction was first used  for the  $\text{SU}(2)$ group  to obtain  rigorous and powerful semiclassical bounds on the quantum partition function \cite{Lieb1973}. This result was  further  generalized  to    arbitrary Lie groups $G$ \cite{Simon1980}.  These bounds then permit a direct extraction of  the DOS  via an inverse Laplace transform of the partition function.  At its core, these bounds  help establish a  general connection between  the DOS of a system and the corresponding classical energy surface. In the following, we will set up the different steps of the methodology required to extract the DOS
for a system with an arbitrary compact symmetry group $G$.  We  first outline   the  construction of generalized coherent states,  then obtain the semiclassical bounds on the quantum partition function and then establish our  result for the DOS.


\subsection{\label{subsec:ClManifold}Construction of Generalized Coherent States and the Classical Phase Space Manifold}
This section largely follows  Chapter 2 of \cite{Perelomov1986}.
Let $\mathcal{H}$ be a Hilbert space equipped with a unitary irreducible representation $T:G\to U(\mathcal{H})$ of a compact Lie Group $G$. Take a fixed vector $\ket{\Psi_0}\in\mathcal{H}$. We define the isotropy subgroup $F$ of $\ket{\Psi_0}$ as the maximal subgroup satisfying the property: 
\begin{equation}
    \forall ~{\rm {f} } \in F: T({\rm f})\ket{\Psi_0} = \exp(i\phi({\rm f}))\ket{\Psi_0}
\end{equation}
Thus $F$ consists of all elements of $G$ whose action on $\ket{\Psi_0}$   induces purely a phase term and leads to physically indistinguishable states.

The  associated classical phase-space manifold $\Gamma$ associated with the pair $(T, \ket{\Psi_0})$ is defined as the coset space $G/F$. For all $\alpha=gF \in \Gamma$, we can define the coherent state associated with $\alpha$ as $\ket{\alpha}:=T(g)\ket{\Psi_0}$, which is unique up to a phase.  The conjunction of the natural measure $\mu$ on $\Gamma$ inherited from $G$ and the coherent states provide  the following  resolution of  the identity operator, 
\begin{equation}
    \mathds{1} = \int_\Gamma d\mu(\alpha)\,\ket{\alpha}\bra{\alpha}.
\end{equation}
 \subsection{Symbols of an Observable and the Berezin Transform}\label{subsec:DiagonalSymbols}
Consider now a    general Hermitian observable $\hat{O}$ on $\mathcal{H}$. Following  the literature on quantization and coherent states \cite{Perelomov1986,Simon1980,Lieb1973,Hall2013},  we  introduce  two real functions termed the  {\it lower and upper symbols }  respectively:  $\Lower{O}$, $\Upper{O}$,  which can be interpreted as classical  variants of  $\hat{O}$:
\begin{align}
    \Lower{O}(\alpha) = \bra{\alpha}\hat{O}\ket{\alpha}
    \,\,\,\text{and}\,\,\,
    \hat{O} = \int_\Gamma d\mu(\alpha)\,\Upper{O}(\alpha)\ket{\alpha}\bra{\alpha}.
\end{align}
For computational and conceptual purposes, it is useful to note that the lower symbol can be written as an integral-transform of the upper symbol.
\begin{equation}\label{eq:Berezin Transform}
    \Lower{O}(\alpha)=\int_\Gamma d\mu(\alpha')\Upper{O}(\alpha')\,|\braket{\alpha'|\alpha}|^2=:\big\{\mathcal{B}\Upper{O}\big\}(\alpha)
\end{equation}
 This transform will be termed {\it the Berezin Tranform}..    
 These  symbols will now be used to derive the semiclassical bounds.

\subsection{Semiclassical Bounds on the Quantum Partition Function}
Let $\hat H$ be the Hamiltonian operator that governs the dynamics of a quantum system. The quantum partition function is  defined as: 
\begin{equation}
    Z_Q(\beta) = \text{Tr}\exp(-\beta {\hat H}) = \int_\Gamma\,d\mu(\alpha)\bra{\alpha}e^{-\beta\hat H}\ket{\alpha}
\end{equation}
Here, the last equality is obtained by evaluating the trace in the coherent state basis. The constructed classical phase space $\Gamma$ and the  upper and lower symbols $(\Upper{H},\Lower{H})$  corresponding to  $\hat H$ allow us to define the  classical partition functions:
\begin{align}
     \Upper{Z}_C(\beta) &:= \int_\Gamma d\mu(\alpha)\,\exp(-\beta \Upper{H}(\alpha))\\
    \Lower{Z}_C(\beta) &= \int_\Gamma d\mu(\alpha)\,\exp(-\beta \Lower{H}(\alpha))
   \end{align}
The inequalities proved in Theorems 2.3 and 2.4 of \cite{Simon1980} describe how the quantum partition function is bounded by the classical partition function. These inequalities have the immediate consequence that:
\begin{equation}
\label{ineq:Simon}
    \Lower{Z}_C(\beta) \leq Z_Q(\beta) \leq \Upper{Z}_C(\beta)
\end{equation}
This inequality is the principal tool for all our subsequent calculations and will be termed the Simon-Lieb inequality. If, in a given system  the overlap $|\braket{\alpha|\alpha'}|^2$ approaches a $\delta$-function on the underlying classical manifold, bounds will saturate. This is  often the case for high dimensional representations of compact groups, notably  SU(2) spins with large $s$. For  reducible representations, the inequality \eqref{ineq:Simon} can still be obtained in  every irreducible sector by restricting the trace in the partition function to states in the irreducible sector.  This permits a direct evaluation of the DOS in the different sectors, as we show in the following subsections.

\subsection{Partition Function to the DOS}
To enable the calculation of the density of states (DOS), we  first assume (without loss of generality) that the ground state energy of the Hamiltonian $\hat H$ and the corresponding  classical energies $\Lower{H}, \Upper{H} $ are  greater than or equal to  $0$. Note that this can always be achieved by adding a sufficiently large constant  to the Hamiltonian. This will result purely in a shift of the DOS, which can then be undone. 
It is easily noted that, under these assumptions, the quantum partition function is  the Laplace transform of the DOS:
\begin{equation} \label{lap}
\text{Tr}e^{-\beta\hat H}=\int_0^\infty dE\,D(E)e^{-\beta E} 
= \mathcal{L}_E[D(E)](\beta)  .  
\end{equation}
Consequently, the inequality in \eqref{ineq:Simon} can be recast as an inequality for the DOS,
\begin{equation}
    \Lower{Z}_C(\beta) \leq \mathcal{L}_E[D(E)](\beta) \leq \Upper{Z}_C(\beta)
\end{equation}
For any given quantum number,  the inverse Laplace transforms of  both LHS and RHS  provide a reasonable approximation to the DOS.
In the limit of large Hilbert space dimension the LHS and the RHS of the inequality approach each other, resulting in tighter semiclassical bounds and more precise estimations of the DOS. 
  Let $h$ denote either $\Upper{H}$ or $\Lower{H}$ of $\hat H$.  The DOS is given by 
\begin{equation}
    D(E)\approx\mathcal{L}^{-1}_\beta[Z_C^h(\beta)](E)
    =\mathcal{L}^{-1}_\beta\left[\int_\Gamma d\mu(\alpha)\,e^{-\beta h(\alpha)}\right](E)\nonumber
\end{equation}
For compact groups $G$,  the order of the inverse transformation and the integration can be interchanged to obtain:
\begin{align} \label{eq:dos}
    D(E)&\approx\int_\Gamma d\mu(\alpha)\,\mathcal{L}^{-1}_\beta[\exp(-\beta h(\alpha))](E)\\
    &=\int_\Gamma d\mu(\alpha)\,\delta(E-h(\alpha))\\
    &\equiv \frac{d U(E)}{dE}
\end{align}
where we have defined the integrated DOS (IDOS).
\begin{equation}\label{eq:IDOSgeneral}
U(E)= \text{Vol}_\Gamma\{\alpha\in\Gamma : h(\alpha)\leq E\}
\end{equation}
In what follows  $\Lower D$, $\Lower U$, and $\Upper D$, $\Upper U$  denote the  DOS and IDOS calculated via lower and upper symbols, respectively.   The inequality \ref{ineq:Simon}  saturates in the limit of high quantum numbers  allowing for the calculation of the precise DOS. Note that the compactness of the group $G$  ensures the finiteness of  the natural measure inherited from $G$ on the classical phase space manifold $\Gamma$.  This feature renders possible a normalized DOS that integrates to unity. The normalization constant will depend only on the Hilbert Space Dimension and the Haar-Measure associated with the group $G$.  For classical phase spaces of low dimension, equation \ref{eq:IDOSgeneral} can be harnessed by using standard numerical integration techniques and can even lead to explicit solutions. For higher dimensional manifolds, this is a multidimensional integral on a constrained manifold and can be computed efficiently via Monte Carlo sampling. Our result \eqref{eq:dos}  indicates that  critical points of the Hamiltonian symbols for a system  should manifest as  irregularities in  the  corresponding   DOS for a general system.  These are non-analyticities and indicate  ESQPTs.   This  correspondence was first  demonstrated  in the  $\text{SU}(2)$ LMG-Model  by \cite{Castanos2006}  and  \cite{Ribeiro_2008}.\\

\subsection{Microcanonical Expectation Values of Observables}
We now discuss how the formal structure of the  DOS  obtained in \eqref{eq:dos}  facilitates the computation of   micro-canonical expectation values.   We use lower symbols $\Lower{O}$ throughout this subsection.  Approximating the  quantum micro-canonical expectation values by their semiclassical counterparts, we obtain
 \begin{align}\label{eq:HellmFeynmExp}
    \langle\hat{O}\rangle&(E) \approx 
    \frac{1}{\Lower{D}(E)}\int_\Gamma d\mu(\alpha)\,\delta(E-\Lower{H}(\alpha))\Lower{O}(\alpha)\nonumber\\
    &=\frac{1}{\Lower{D}(E)}\frac{d}{d\varepsilon}\int_\Gamma d\mu(\alpha)\,\theta(\varepsilon-\Lower{H}(\alpha))\Lower{O}(\alpha)\biggr\rvert_{\varepsilon= E}
\end{align}
These calculations can be made with the upper symbols analogously. 

For general observables, these integrals can be evaluated.  For  observables  given by the derivative of the Hamiltonian  w.r.t. some parameter  $c$ (e.g. field strength, coupling constant, etc.),   a Hellman-Feynman like scheme simplifies the calculation.  We assume that the Hamiltonian  is given by ${\hat H}(c_0)$ and  the observable $\hat{O} = \frac{\partial \hat H(c)}{\partial c}\rvert_{c=c_0}$. 
To simplify the notation, we write $\Lower{H(c)}(\alpha)=\Lower{H}(c,\alpha)$.  We find that the microcanonical expectation value can expressed in terms of the  DOS  $\Lower D$ and the IDOS $\Lower{U}$
\begin{align}\label{observables}
    \langle\hat{O}\rangle(E) &\approx \frac{1}{\Lower{D}_{c_0}(E)}\int_\Gamma d\mu(\alpha)\delta(E-\Lower{H}(c_0, \alpha))\,\frac{\partial\Lower{H}(c,\alpha)}{\partial c}\biggr\rvert_{c=c_0}\nonumber\\
    &=-\frac{1}{\Lower{D}_{c_0}(E)}\frac{\partial}{\partial c}\int_\Gamma d\mu(\alpha)\,\theta(E-\Lower{H}(c,\alpha))\biggr\rvert_{c=c_0}\nonumber\\
    &=-\frac{1}{\Lower{D}_{c_0}(E)}\frac{\partial}{\partial c}\Lower U_c(E)\biggr\rvert_{c=c_0}
\end{align}
This  result  generalizes the  approach of  \cite{Ribeiro_2008, Brandes2013}  for spin observables.

\subsection{Bounds on the Ground State Energy\label{subseq:GSbounds}}

For compact symmetry groups $G$,  the symbols $\Lower{H}$ and $\Upper{H}$  have well-defined minimum and maximum values\cite{Boundedness}. We define the semiclassical ground state energies as
\begin{equation}\label{eq:minmaxenergy}
    \Lower{E}_0:=\min_{\alpha\in\Gamma}\Lower{H}(\alpha)\,\,\,\text{and}\,\,\,\Upper{E}_0:=\min_{\alpha\in\Gamma}\Upper{H}(\alpha).
\end{equation}
Next, we define semiclassical and quantum free energies.
\begin{align}
    F_Q(\beta) &:= -\beta^{-1}\log Z_Q(\beta) \\
    \Lower F_C(\beta) &:= -\beta^{-1}\log \Lower Z_C(\beta)
\end{align}
The free energies for the upper symbol are defined analogously. The Simon-Lieb inequality \eqref{ineq:Simon}  translates to bounds for the quantum free energy:
\begin{equation}\label{ineq:freeenergies}
    \Upper F_C(\beta)\leq F_Q(\beta) \leq \Lower F_C(\beta)
\end{equation}
In both the semi-classical and quantum cases, the derivative of the free energy w.r.t. $\beta$ yields the canonical expectation value for energy:
\begin{align}
    \partial_\beta F_Q(\beta) &= \frac{1}{Z_Q(\beta)}\text{Tr}[\hat{H}e^{-\beta \hat{H}}]=\langle\hat H\rangle_Q(\beta) \\
    \partial_\beta \Lower F_C(\beta) &= \frac{1}{\Lower Z_C(\beta)}\int_\Gamma d\mu(\alpha) \Lower H(\alpha)e^{-\beta \Lower H(\alpha)} = \langle \Lower H\rangle_C(\beta)\nonumber
\end{align}
Taking the  $\beta \to \infty$  limit in the above equations results in explicit bounds for the quantum ground-state energy $E_0=\lim_{\beta\to\infty}\partial_\beta F_Q(\beta)$. 
As detailed in Appendix \ref{App:proofGS} we find:
\begin{equation}\label{ineq:GSbounds}
    \Upper E_0 \leq E_0 \leq \Lower E_0
\end{equation}
To summarize, the Simon-Lieb inequality provides a direct pathway to obtain the density of states, ground state energies, and microcanonical expectation values in every high dimensional irreducible sector of a quantum manybody system.

\section{Results for the Lipkin Meshkov Glick Model\label{sec:LMG-Model}}
As a proof of concept, we now apply  our methodology  to the well-known Lipkin Meshkov Glick model to study both QPTs and ESQPTs within this model. We choose this model as our test case for two principal reasons: (i) the all-to-all couplings in this model ensure that the Hamiltonian is block diagonal with respect to the individual spin sectors, and (ii) the existence of analytical results \cite{Ribeiro_2008} and extensive recent numerical results \cite{corps2024} for the DOS. The Hamiltonian describing $N$ all-to-all interacting spin-$1/2$ systems in a magnetic field $\kappa$ reads:
\begin{equation}
    \hat H = -\frac{1}{N}(\gamma_xS_x^2+\gamma_yS_y^2)-\kappa S_z
\end{equation}
Here $\vec S=\sum_{i=1}^N\frac{\vec\sigma_i}{2}$ where $\vec\sigma_i$ are the Pauli spin matrices. For convenience, we drop the hat symbol on the spin operators. Due to the all-to-all interaction, we have $[\hat H, \vec S^2]=0$. The Hilbert space decomposes into irreducible representations of $\text{SU}(2)$ via Clebsch-Gordon. The Hamiltonian is block-diagonal with respect to these irreducible sectors.

The Hamiltonian  is invariant under a  $\pi/2$ rotation around the z-Axis ($e^{-i\frac{\pi}{2}S_z}$)  which maps $S_x\mapsto S_y$ and $S_y\mapsto -S_x$. This ensures that the phase diagram (see Figure \ref{Plots:DOSLMGZones}(a)) is symmetric under $\gamma_x\leftrightarrow\gamma_y$.  In the following, we restrict our discussion to the case $\kappa>0$  as the $\kappa <0$ case  can be accessed via a $\pi$ rotation  of the Hamiltonian around the x-Axis $e^{-i\pi S_x}$  which maps $\kappa \to -\kappa$. 

We will focus on the regime $\gamma_x\geq-\gamma_y$, where the ground state lies near the higher spin sector (the ferromagnetic interaction dominates). There are three well known quantum phases in this regime. For $\kappa>\max(\gamma_x,\gamma_y)=:\gamma_{\max}$, the ground state is unique, and the spins are aligned along the field-direction. The ground state is symmetric under the $\mathbb{Z}_2$ transformation given by a $\pi$ rotation around the z-Axis ($e^{-i\pi S_z}$). For $\kappa<\gamma_{\max}$ and $\gamma_x\neq\gamma_y$, the spins align (positively or negatively) along the axis with the maximal coupling constant, spontaneously breaking the $\mathbb{Z}_2$ symmetry. If $\gamma_x=\gamma_y$ the ground state breaks a $\text{U}(1)$ rotational symmetry around the z-Axis, yielding a new quantum phase.  We now  set up the coherent state manifold for this problem.

\subsection{$\text{SU}(2)$ Coherent States and Classical Phase Space}\label{subsec:SU(2)CoherentStates}
\begin{table*}[pt]
\centering

\caption{\label{tab:symbols}%
Symbols of some elementary Observables
}
\begin{tabular}{|c|c|c|}
\colrule
$\hat{O}$&
$\Lower{O}$&
$\Upper{O}$\\
\colrule
$S_z$ & $s\cos\theta$ & $(s+1)\cos\theta$\\
$S_x$ & $s\sin\theta\cos\varphi$ & $(s+1)\sin\theta\cos\varphi$ \\
$S_y$ & $s\sin\theta\sin\varphi$ & $(s+1)\sin\theta\sin\varphi$\\
$S_x^2$ & $s(s-\frac{1}{2})(\sin\theta\cos\varphi)^2+\frac{s}{2}$ & $(s+1)(s+\frac{3}{2})(\sin\theta\cos\varphi)^2-\frac{s+1}{2}$\\
$S_y^2$ & $s(s-\frac{1}{2})(\sin\theta\sin\varphi)^2+\frac{s}{2}$ & $(s+1)(s+\frac{3}{2})(\sin\theta\sin\varphi)^2-\frac{s+1}{2}$\\
\colrule
\end{tabular}
\end{table*}
Let $T_s:\text{SU}(2)\to \text{U}(\mathcal{H}_s)$ denote the spin-$s$ representation of $\text{SU}(2)$. We apply the general construction discussed in \ref{subsec:ClManifold} to define $\text{SU}(2)$ coherent states and the classical phase space manifold associated with  $\text{SU}(2)$. Following the choice in the literature, we choose $\ket{\psi_0} = \ket{s,s}$ which can be intuitively thought of as the spin pointing to the north-pole along the quantization axis (which we choose to be the z-axis). We notice that the isotropy subgroup (subgroup whose action on $\ket{\psi_0}$ is physically indistinguishable) consists of all rotations around the z-Axis, i.e. $F=U(1)$. We get $\Gamma = \text{SU}(2)/\text{U}(1)\cong \text{SO}(3)/\text{SO}(2)\cong S^2$. This is in accordance with the representation of   the quantum spin as a point on the two-sphere $S^2$. Let $(\varphi, \theta) =: \Omega\in S^2$ be a point on the two-sphere. Writing out a general $\text{SU}(2)$ rotation in terms of the Euler Angles $R(\theta,\varphi,\vartheta)=e^{-i\varphi S_z}e^{-i\theta S_y}e^{-i\vartheta S_z}$, and dividing out the Isotropy Subgroup $e^{-i\vartheta S_z}$, we obtain the spin coherent state.
\begin{align}\label{eq:SU(2)CohState}
    \ket{\Omega} = \ket{\theta,\varphi}= e^{-i\varphi S_z}e^{-i\theta S_y}\ket{s,s}
\end{align}
We note that alternative parametrizations of the coherent states exist. We obtain the following resolution of identity on $\mathcal{H}_s$: 
\begin{equation}
    \mathds{1} = \frac{2s+1}{4\pi}\int_{S^2}d\Omega\,\ket{\Omega}\bra{\Omega}
\end{equation}
The absolute value squared of the overlap can be calculated to be: 
\begin{equation}
    |\braket{\Omega|\Omega'}|^2=\Big[\cos\frac{\mathbf{\Theta}}{2}\Big]^{4s}
\end{equation}
Where $\mathbf{\Theta}$ denotes the angle between the points $\Omega$ and $\Omega'$ on the sphere. It is easily noticed that for high $s$ this term asymptotically approaches a $\delta$ function on the sphere.  
\begin{equation}\label{eq:SpinOverlapDeltaAsymp}
    |\braket{\Omega|\Omega'}|^2\simeq \frac{\delta^{S^2}(\Omega-\Omega')}{2s+1}
\end{equation}
This means that in the high $s$ limit, the Berezin Transform approaches the identity, resulting in the saturation of the Simon-Lieb inequality \ref{ineq:Simon}.

\subsection{Lower \& Upper Symbols and their calculation via the Berezin Transform}
Next, we present the lower and upper symbols of some elementary observables as they can be found in \cite{Lieb1973} and \cite{Perelomov1986}. The calculation of the lower symbols is a straightforward computation of expectation values. Alternatively, and more generally, the lower symbol of any spin observable can be calculated via a generating function discussed in \cite{Arecchi1972}. The calculation of upper symbols inevitably becomes more involved; the most general pathway involves the Berezin Transform, as introduced in equation \eqref{eq:Berezin Transform}. 
\begin{align}
    \mathcal{B}\Upper{O}(\Omega)&=\frac{2s+1}{4\pi}\int_{S^2}d\Omega'\Upper{O}(\Omega')|\braket{\Omega'|\Omega}|^2\\
    &=\Lower{O}(\Omega)
\end{align}
Thus, we need to invert the Berezin Transform to obtain the upper symbol. To this end, we prove in Appendix \ref{app: Berezin Transform Diagonalization} that the Berezin Transform is diagonalized by spherical harmonics with the following eigenvalues:
\begin{align}\label{eq:BerezinEigenvalues}
    \mathcal{B}Y_{lm}&=\frac{(2s)!(2s+1)!}{(2s-l)!(2s+l+1)!}Y_{lm},\, 0\leq l\leq 2s\\
    \intertext{The inverse becomes straight-forward:}
    \mathcal{B}^{-1}Y_{lm}&=\frac{(2s-l)!(2s+l+1)!}{(2s)!(2s+1)!}Y_{lm},\, 0\leq l\leq 2s
\end{align}
With the help of this identity, one can expand each lower symbol into a sum of spherical harmonics to compute the upper symbol. Using  the symbols listed in Table \ref{tab:symbols}, we     directly evaluate   the lower and upper symbols of the LMG-Hamiltonian in an arbitrary spin-sector.
\begin{widetext}
\begin{align}\label{eq:hamsymb}
    \Lower H(\Omega) &=-\frac{1}{N}\{\gamma_x(s(s-\frac{1}{2})(\sin\theta\cos\varphi)^2+\frac{s}{2})+\gamma_y (s(s-\frac{1}{2})(\sin\theta\sin\varphi)^2+\frac{s}{2})\}-\kappa s\cos\theta\\
    \Upper H(\Omega) &=-\{\frac{\gamma_x}{N}((s+\frac{1}{2})(s+\frac{3}{2})(\sin\theta\cos\varphi)^2-\frac{s+1}{2})\nonumber
    +\frac{\gamma_y}{N}((s+\frac{1}{2})(s+\frac{3}{2})(\sin\theta\sin\varphi)^2\nonumber-\frac{s+1}{2})\}-\kappa(s+1)\cos\theta\nonumber
\end{align}
\end{widetext}
These symbols can now be used to derive the Simon-Lieb bounds (\ref{ineq:Simon}).

 \subsection{Ground State Energy and QPT \label{sec:GSLGM}}
\begin{figure}[h]
\centering
\begin{overpic}[width=0.48\linewidth]{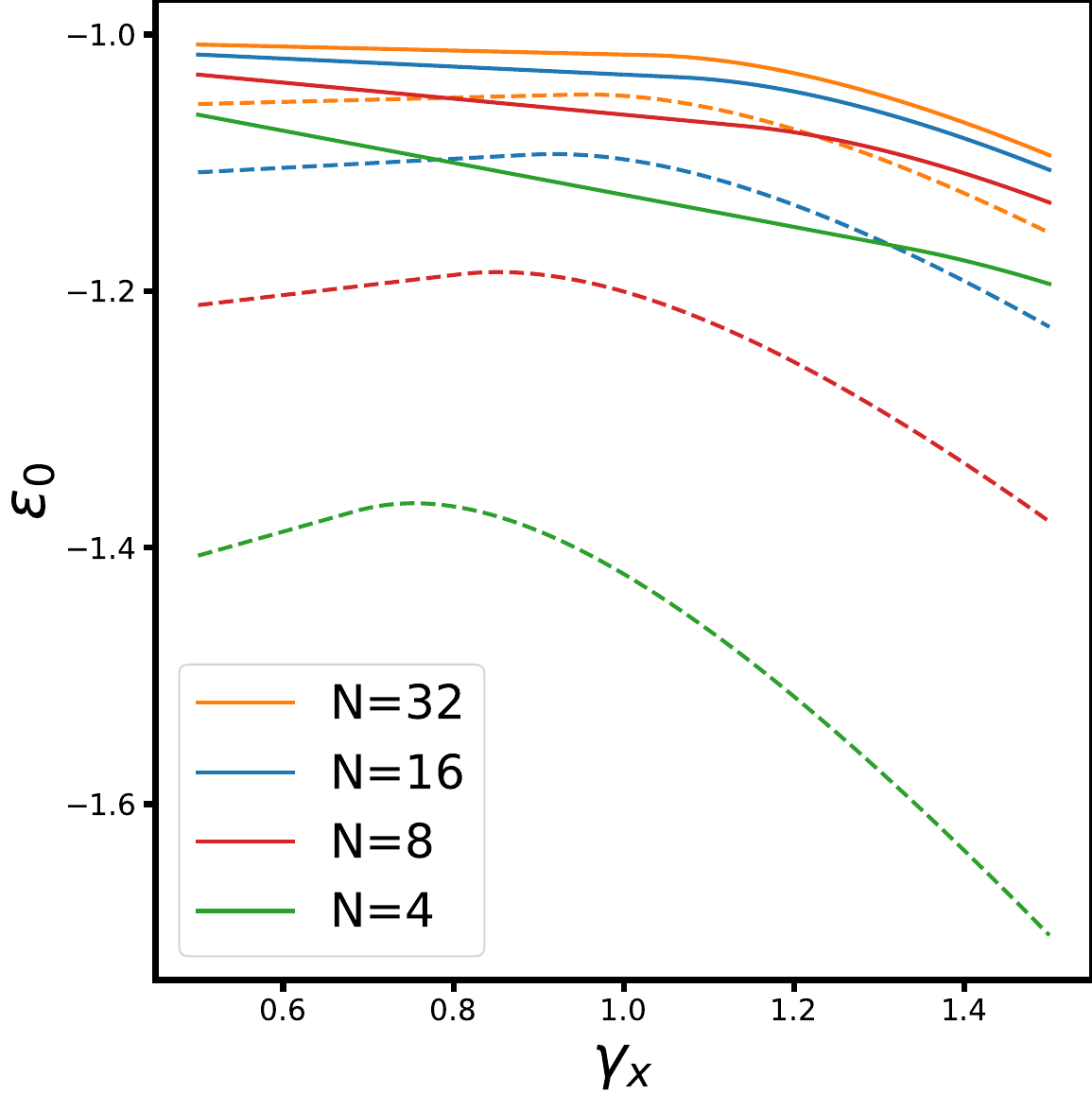}
    \put(5,105){\large (a)}
\end{overpic}
\begin{overpic}[width=0.48\linewidth]{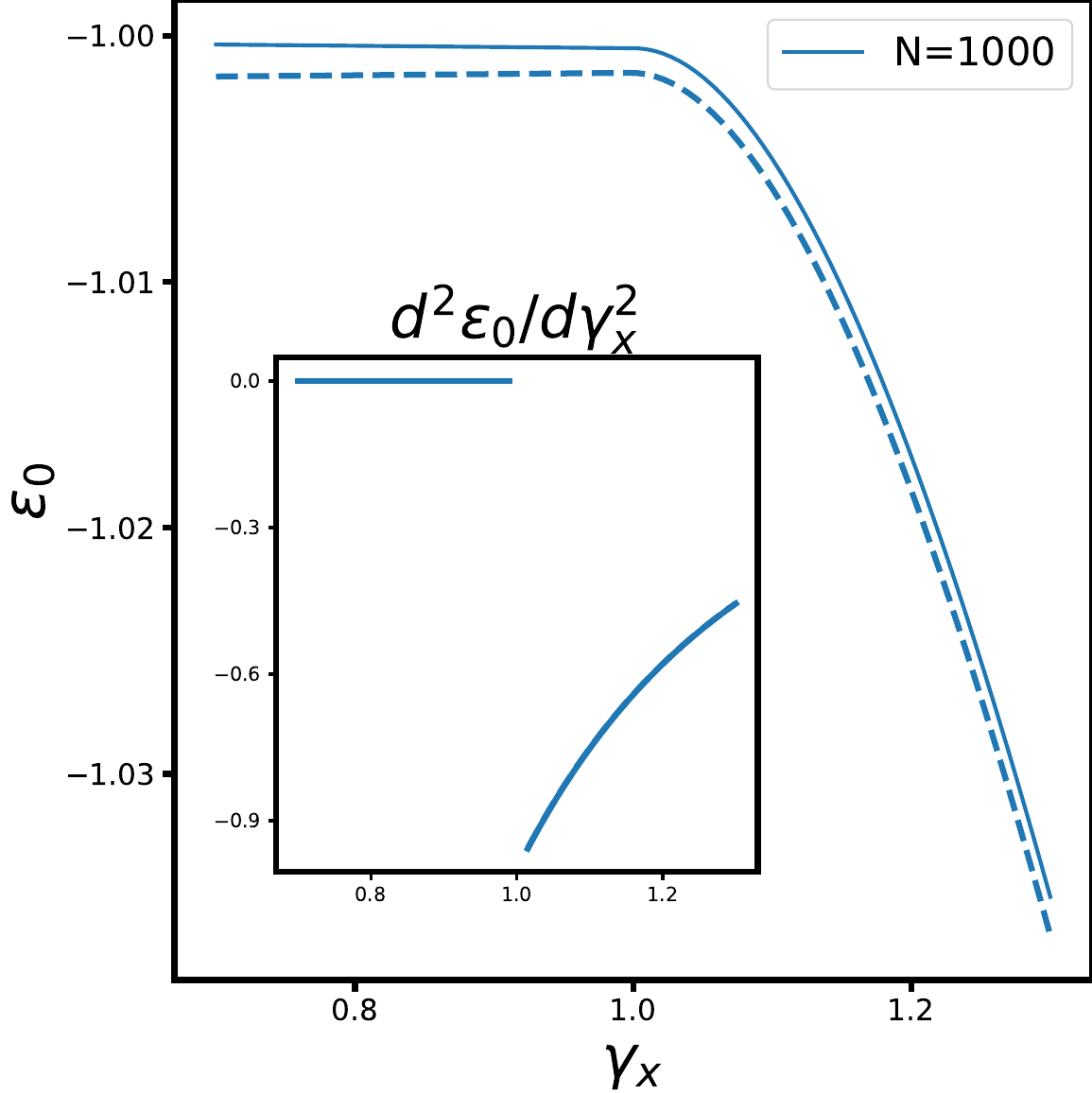}
    \put(5,105){\large (b)}
\end{overpic}
\caption{Lower and upper bounds for the ground state energy density $\varepsilon_0(\gamma_x) = 2E_0(\gamma_x)/N$ as a function of $\gamma_x$, calculated via  \ref{ineq:GSbounds}. Dashed (solid) lines are the lower (upper) bounds for a given particle number. It becomes clear that for increasing particle number two bounds approach each other. The second derivative of the parameter-dependent ground-state energy density in the high particle number demonstrates the second-order QPT.}
\label{fig:QPT}
\end{figure}

 We demonstrate the power  of the method discussed in subsection \ref{subseq:GSbounds} to  explicitly bound the ground state energy $E_0  $ of the system. Here we choose coupling parameters that satisfy $\gamma_x, \gamma_y \geq 0$. This ensures that the ground state lies within the highest spin sector, as we have  proven in Appendix \ref{App:Proof LMG GS}. Our results for the  lower and upper bounds for $\varepsilon_0(\gamma_x) = 2E_0(\gamma_x)/N$ ($\gamma_y=0$,  and overall scaling $\kappa=1$) and varying system size $N$ are shown in Fig.~\ref{fig:QPT}. These values are obtained by numerically calculating the minima as defined in equation \eqref{eq:minmaxenergy} for the compact phase space $\Gamma = S^2$. For this compact case, a numerical gradient descent algorithm is stable, sufficient and fast.\\
 \\
As we see in Fig.~\ref{fig:QPT}(a),  the bounds approach each other with increasing $N$ and yielding converged results for the ground state energy for  $N=1000$.  We also note the discontinuity in
the second order derviative of the bounds as $\gamma_x \to 1$ confirming  the existence of a QPT at $\gamma_x\to 1$ (inset of Fig.~\ref{fig:QPT}(b) ).

\subsection{DOS in the Highest Spin Sector and ESQPT}
\begin{center}
\begin{figure*}[t]

\centering
\begin{overpic}[width=0.24\linewidth]{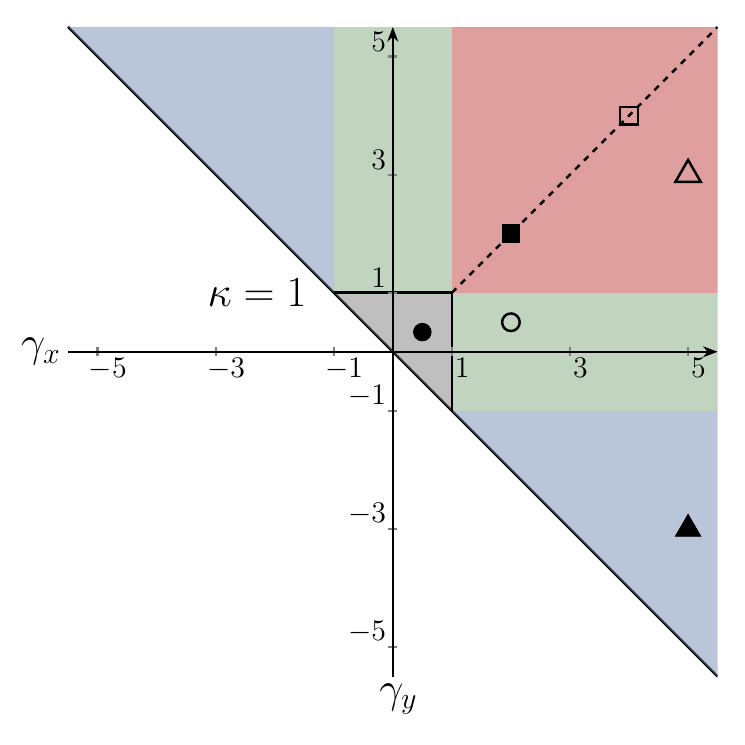}
    \put(5,100){\large (a)}
\end{overpic}
\begin{overpic}[width=0.24\linewidth]{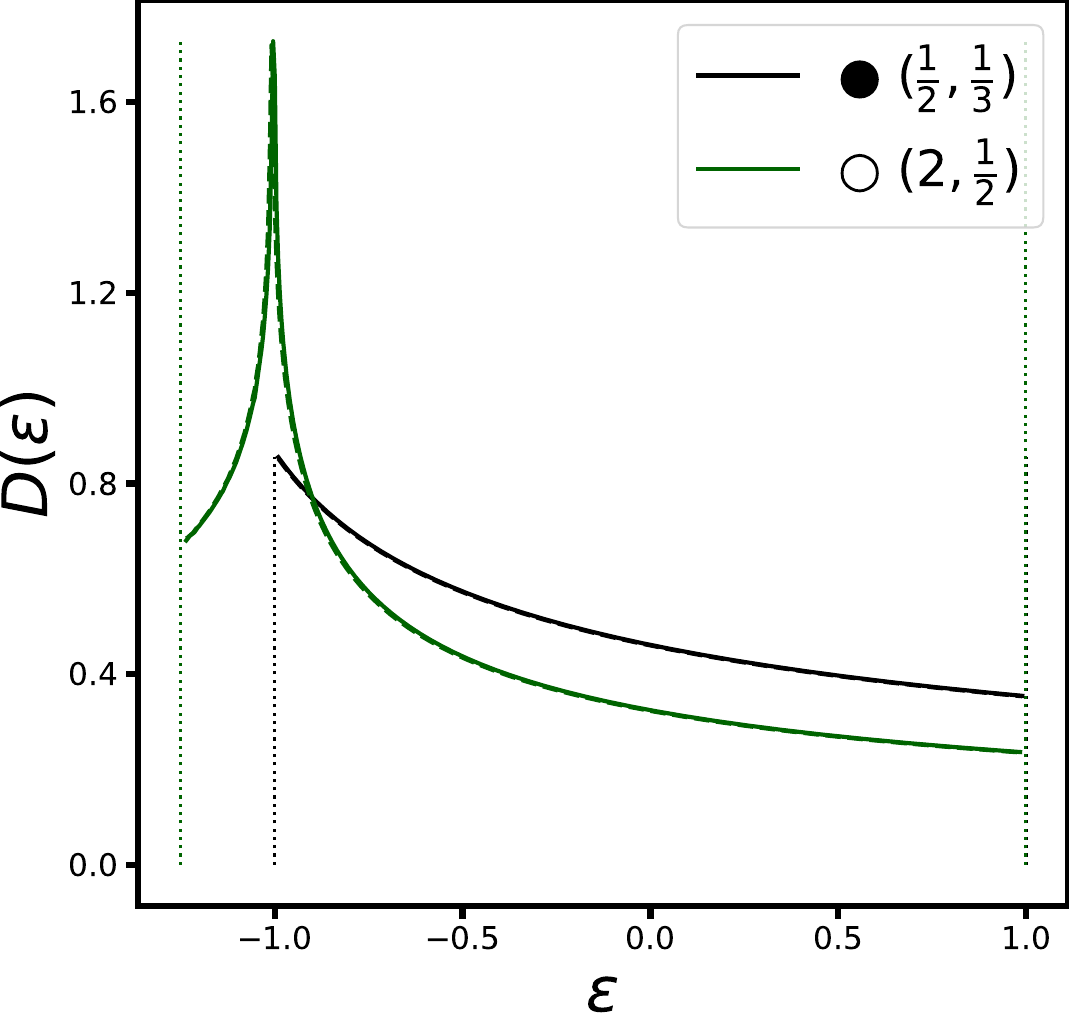}
    \put(5,100){\large (b)}
\end{overpic}
\begin{overpic}[width=0.24\linewidth]{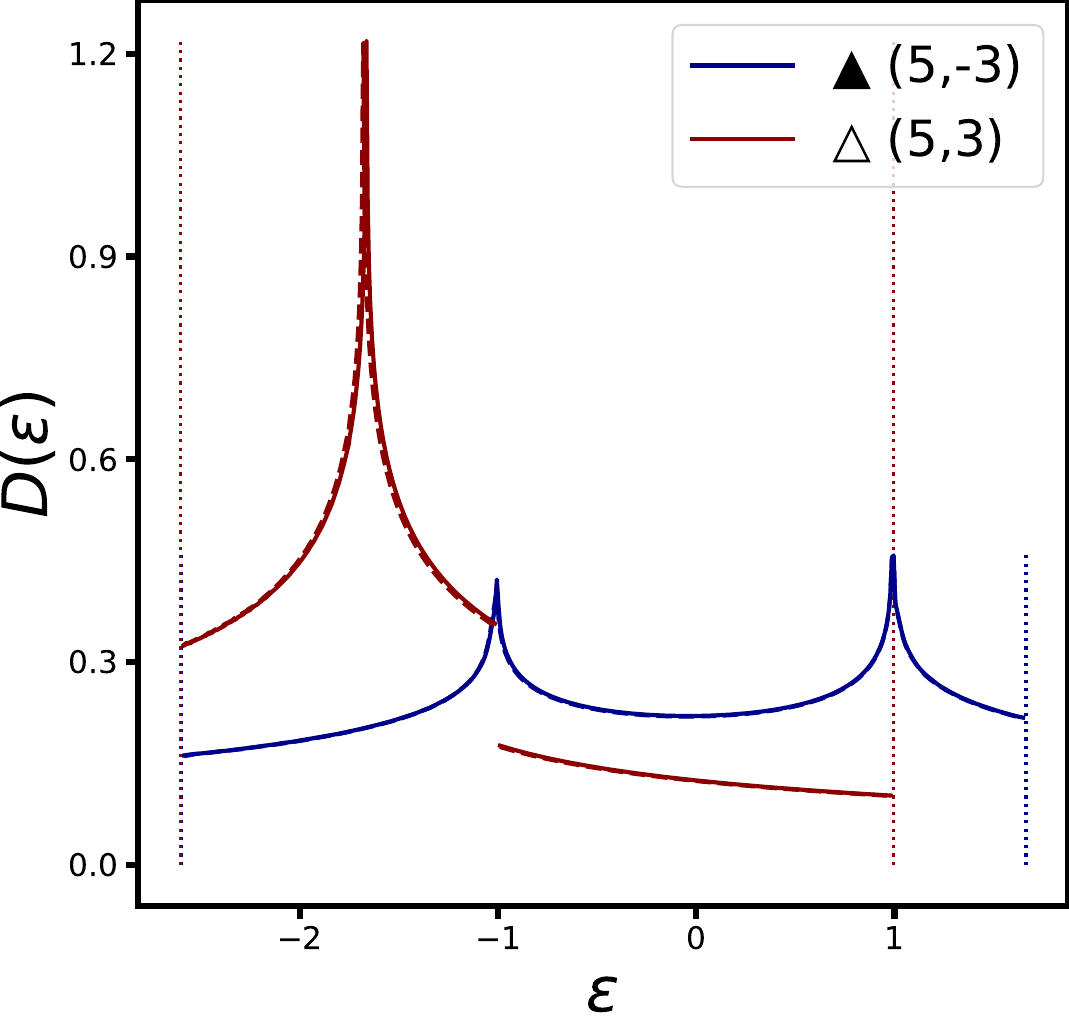}
    \put(5,100){\large (c)}
\end{overpic}
\begin{overpic}[width=0.24\linewidth]{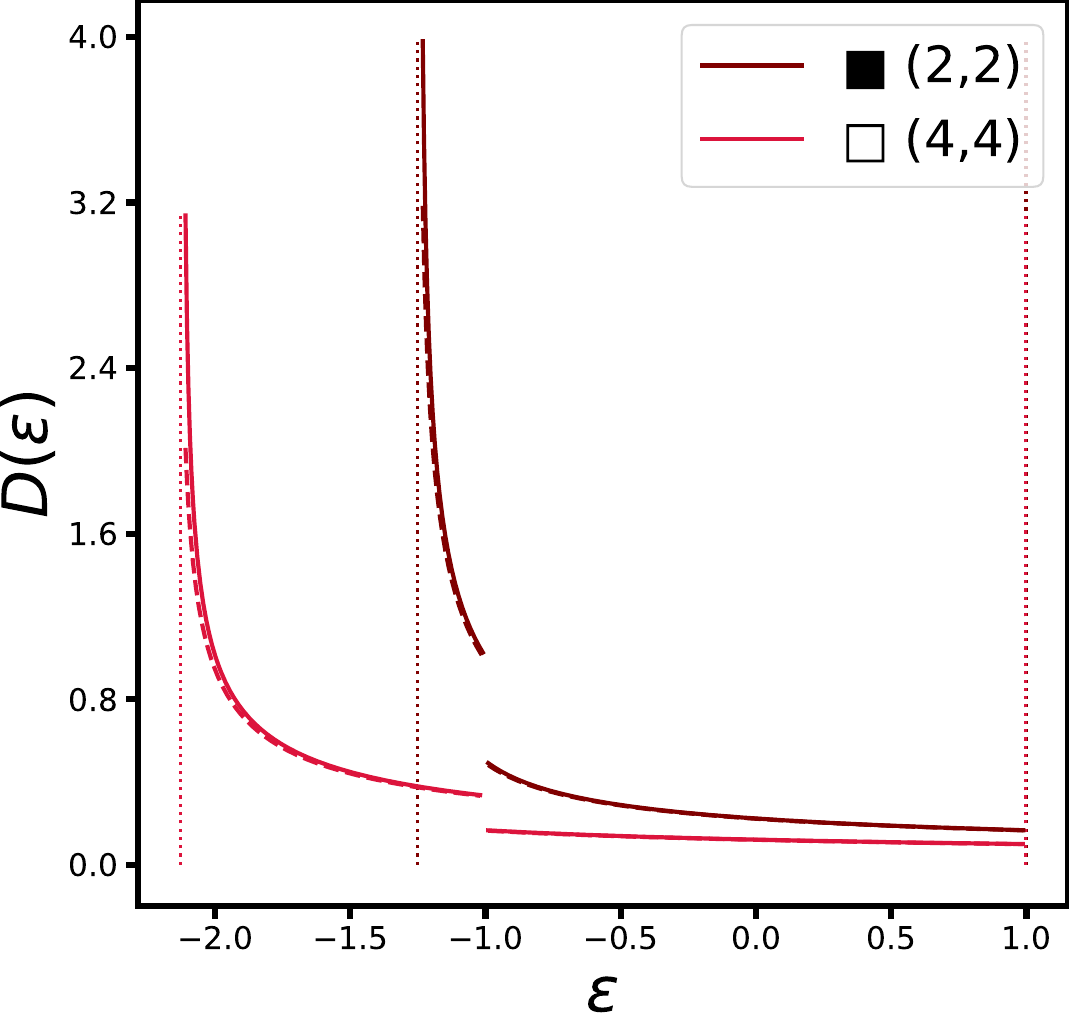}
    \put(5,100){\large (d)}
\end{overpic}
\caption{(a)The ESQP-Diagram of the LMG-Model. For the DOS calculations we have fixed $N=1000$, $\kappa=1$. In this high $N$ limit, we distinguish between four different phases according to the DOS behavior. Black: Analytical, Green: One singularity, Blue: Two singularities, Red: One singularity and one discontinuity. (b)(c)(d) show the DOS for various parameter pairs $(\gamma_x,\gamma_y)$ marked on the phase diagram. The energy axis show the scaled energy $\varepsilon=2E/N$. The dashed curves show DOS calculated via the upper symbol and solid curves show the DOS calculated via the lower symbol. In this high $N$ limit these have converged as expected and are barely distinguishable. The vertical dotted lines indicate the ground state energy and the maximal possible energy for each parameter pair. The DOS plotted in (d) highlight the distinct Quantum Phase along $\gamma_x=\gamma_y$. For $\gamma_x=\gamma_y$ the ground state becomes a manifold with the associated $\text{U}(1)$ rotational symmetry. This manifests itself as a singularity located at the ground state. It is reasonable to assume that the singularity visible for different parameter pairs is a similar manifold of states accessible only at higher energies.}
\label{Plots:DOSLMGZones}
\end{figure*}
\end{center}

Inserting the Hamiltonian symbols  \eqref{eq:hamsymb}  (with $s=N/2$) into the general inequality \ref{ineq:Simon} we obtain:
\begin{align}\label{eq:bound}
    \frac{N+1}{4\pi}\int_{S^2}d\Omega\, e^{-\beta \Lower{H}(\Omega)}\leq \text{Tr}e^{-\beta\hat H}\\ \leq\frac{N+1}{4\pi}\int_{S^2}d\Omega\, e^{-\beta \Upper{H}(\Omega)} \nonumber
\end{align}
Dividing  all sides of this inequality by $(N+1)$ enforces $\lim_{E\to\infty} U(E) = 1$. In other words, the DOS $D(E)$ is normalized such that its integral over the whole spectrum in the given irreducible sector equals $1$ (regardless of particle number). The DOS is obtained through \eqref{eq:dos}. 
\begin{align}
\label{eq:LMG DOS}
    \Lower{U}(E) &= Vol_{S^2}\{\Omega\in S^2:\Lower{H}(\Omega)\leq E\} \\
    \Lower D(E) &= \frac{d}{dE}\Lower U(E)
\end{align}
and, analogously, for $\Upper H$.  Our results for the DOS are summarised in Fig.~\ref{Plots:DOSLMGZones}. 
In contrast to the two ground state  phases for  $\max(\gamma_x,\gamma_y)<\kappa$ and $\max(\gamma_x,\gamma_y)>\kappa$,  the DOS shown in Fig.~\ref{Plots:DOSLMGZones} delineates four different regions  characterized by the number and types  of non-analyticities  in the DOS. These non-analyticities  can be discontinuities (see Figure \ref{Plots:DOSLMGZones}(c),(d) or singularities (see Figure \ref{Plots:DOSLMGZones}(b),(c),(d)) and signal ESQPTs at these energies. Note that there are no true singularities in the finite $N$ case, but we expect that these cusps seen in the finite $N$ DOS will become singularities in the thermodynamic limit.

Previous calculations of the DOS in related models make explicit use of the structure of the Hamiltonian, its eigenstates, and the specific sector under consideration. In \cite{Ribeiro_2008}, the authors  rely on the counting of  zeros of  majorana polynomials associated with the eigenstates of the LMG model to obtain the DOS, while  \cite{Brandes2013} uses the explicit Dicke Basis for the highest spin sector to evaluate a necessary spin trace. It is unclear how these approaches would generalize to other spin sectors or more complicated underlying algebras. Our general approach through equation \ref{eq:LMG DOS}  obviates the need for such complex structures.  We now present our results for the DOS in the different spin sectors.

\subsection{\label{subsec:LGMBeyond}DOS beyond the highest spin sector}
\begin{figure}[h]
 \includegraphics[width=0.45\textwidth]{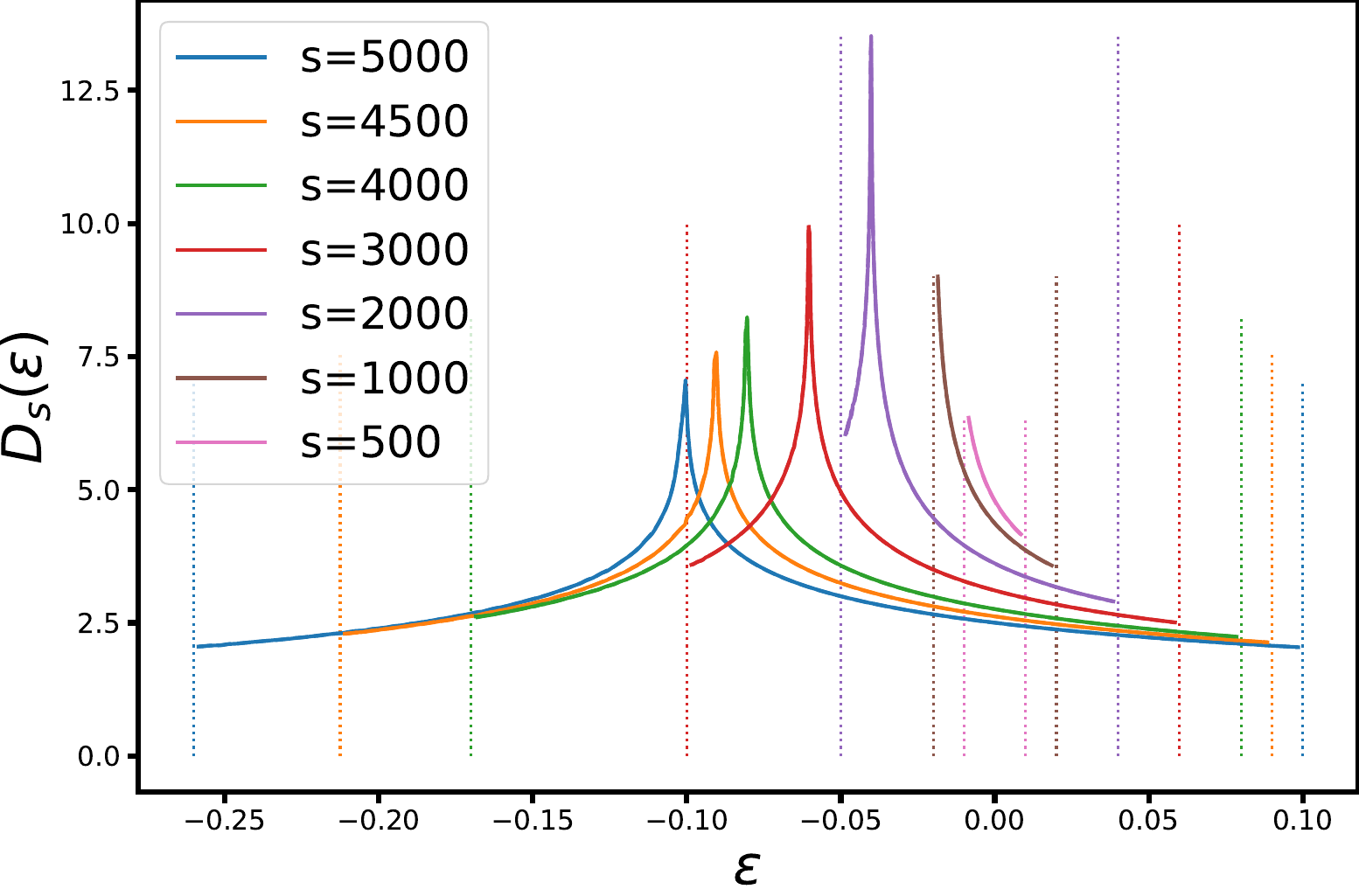}
 \caption{DOS of different spin sectors. The energy axis has the same scaling $\varepsilon=2E/N$ for every sector. The DOS states of each sector is normed by dimension of the highest spin sector.This calculation is in perfect agreement with the numerical results in \cite{corps2024}.}
\label{Plots:SectorsDOS} 
\end{figure}

Applying the Simon-Lieb  inequality (\ref{ineq:Simon})  to lower spin sectors, we can obtain  the  DOS in a straightforward manner.
Though $D(E)$ is normalized to one independently in each spin sector,  to facilitate a comparison  between the number of states in different spin sectors, we normalize the density of states (DOS) of each sector by the dimension of the highest spin sector $s=N/2$. 
\begin{equation}
    \int_{-\infty}^\infty D_s(E)\,dE = \frac{2s+1}{N+1} 
\end{equation}
Where $D_s$ denotes the DOS of the spin-sector $s$.  To facilitate a direct comparison with   recent numerical estimations of the DOS in \cite{corps2024}, we now  fix $(\gamma_x,\gamma_y,\kappa)=(0.5,0,0.1)$ and $N=10^4$.  Note that the highest-spin sector of the chosen parameters  corresponds to the green phase  of  Fig.~\ref{Plots:DOSLMGZones} with one singularity. Our results for $D_s$  for different spin sectors $s$ are shown in Fig.~\ref{Plots:SectorsDOS} and are in perfect  quantitative agreement with the numerical results in \cite{corps2024}.   Note that the bandwidth of the DOS decreases with decreasing $s$. We find that the non-analytic peaks  disappear for $s\le 1000$. 
\\
Both these aspects can be simply explained  from our coherent state perspective. First,  as shown in Table \ref{tab:symbols}, the symbols corresponding to $S_x^2, S_y^2$ scale as $s^2$
while  that of  $S_z\propto s$. Consequently,  lowering the spin sector  can alternately also be viewed as an  effective  quadratic reduction  of  $\gamma_x$ and $\gamma_y$  and a linear reduction of $\kappa$.  A naive calculation following this would suggest that the singularity in the DOS will vanish when $2s/N =\kappa/\gamma_x$ holds i.e.,  $s \le 1000$  as the effective couplings now correspond to Phase I, which as shown in Fig.~\ref{Plots:SectorsDOS} does not possess any singularities. To summarise,  our approach provides for an easy  estimation of the DOS in different spin sectors.

\subsection{\label{sec:Microcan}Microcanonical Observable Expectation Values in the Highest Spin Sector}
\begin{figure}[h]
\centering
\begin{overpic}[width=0.48\linewidth]{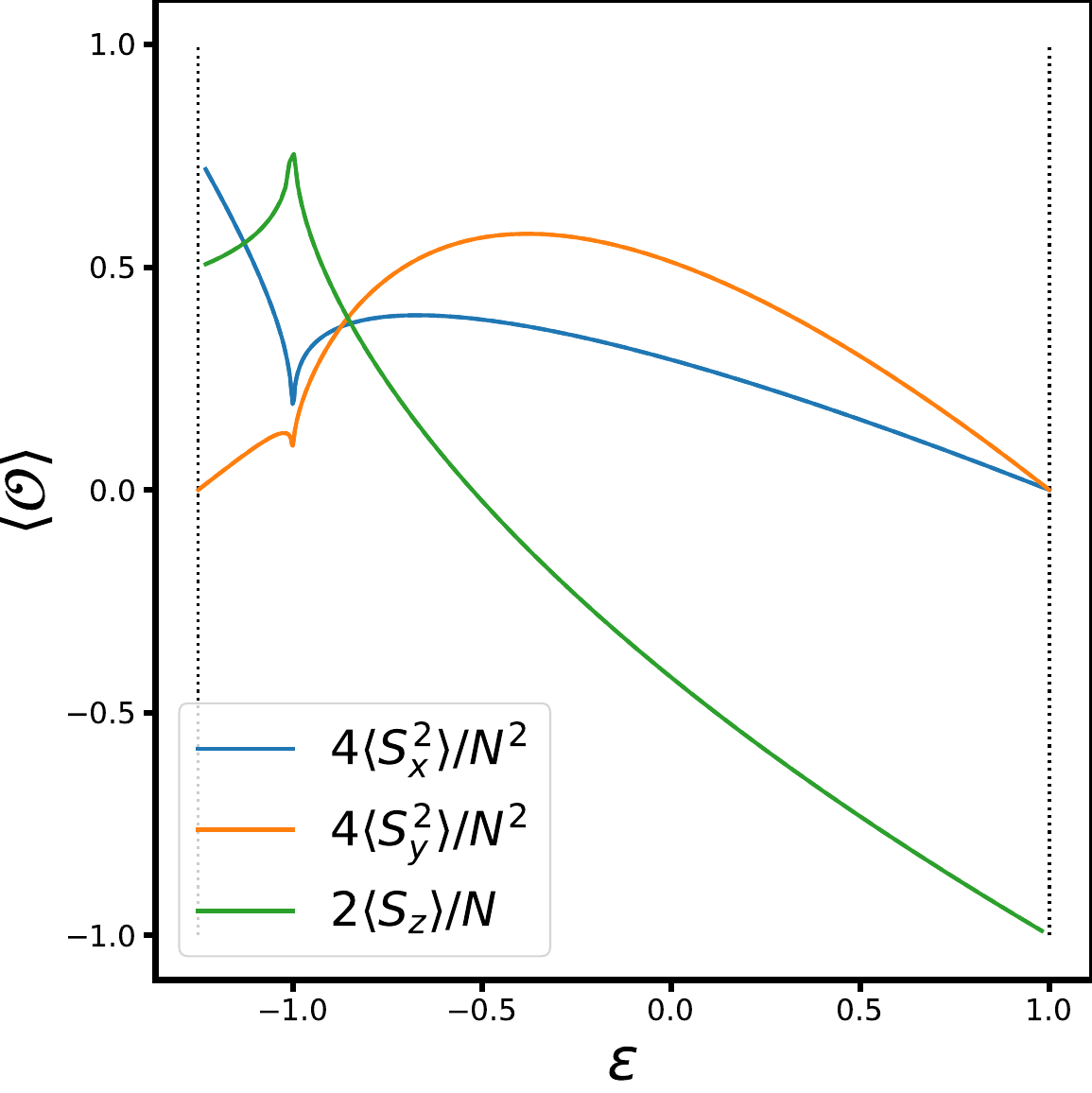}
    \put(5,105){\large (a)}
\end{overpic}
\begin{overpic}[width=0.48\linewidth]{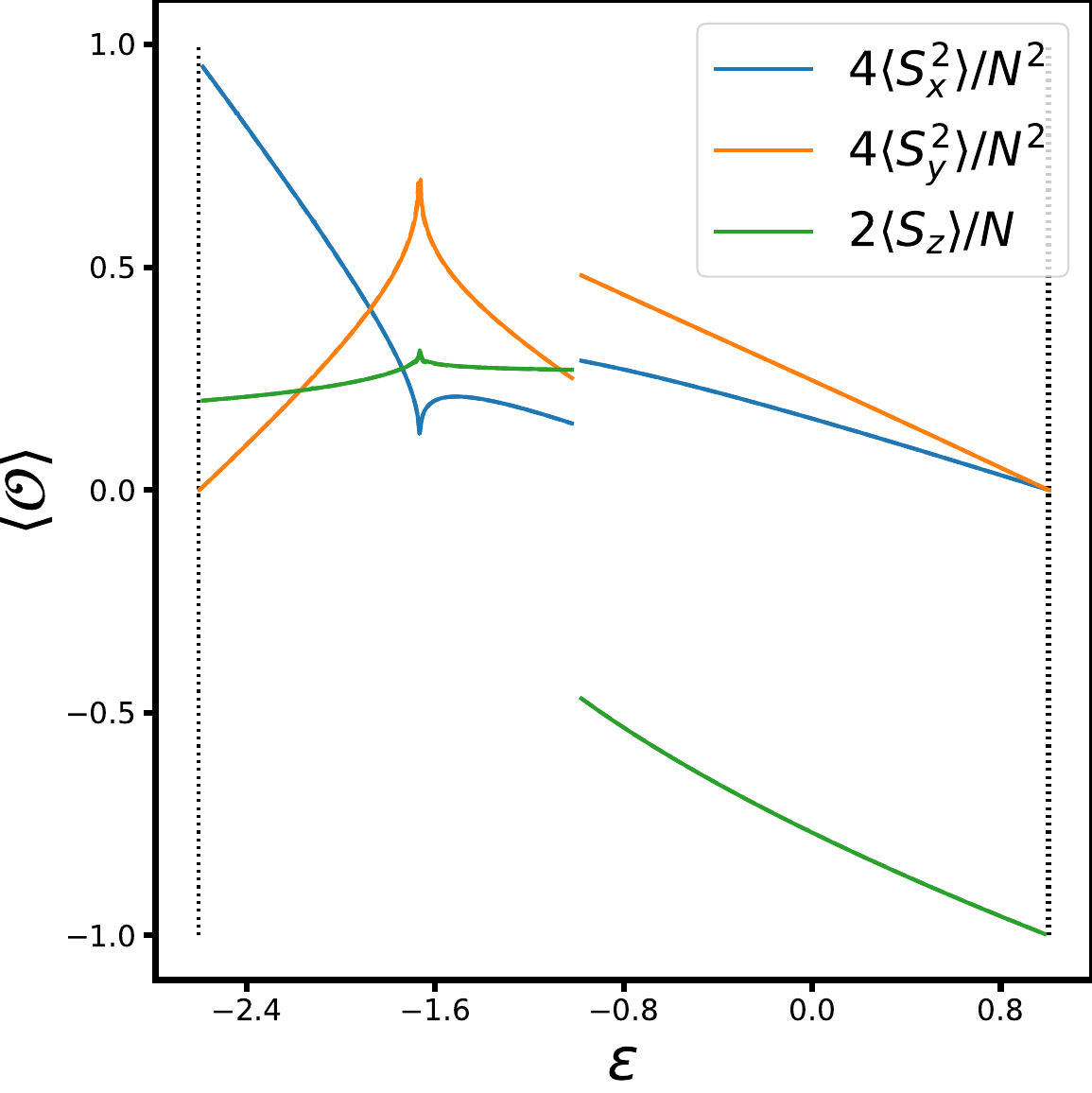}
    \put(5,105){\large (b)}
\end{overpic}
\caption{Microcanonical expectation values for spin observables calculated for fixed parameters $N=1000$ and $\kappa=1$: (a) for green phase parameters $(\gamma_x,\gamma_y)=(2,1/2)$ and (b) for red phase  parameters $(\gamma_x,\gamma_y)=(5,-3)$. The ground-state and maximal-state energies are marked by dashed lines (calculated by both symbols). The observable expectation values are set to $0$ outside these energies. Note that the non-analyticities in the expectation values correspond to those observed in the DOS of each phase.
  These calculations are in perfect agreement with those in ~\cite{Ribeiro_2008}.} 
  
\label{Plots:SpinObservables}
\end{figure}

We now compute the  microcanonical expectation values of spin observables  in the highest spin sector $s= \frac N 2$. The relevant operators can be recast as
\begin{align}
    S_x^2 = -N\frac{\partial {\hat H}}{\partial\gamma_x},\,\,S_y^2 = -N\frac{\partial {\hat H}}{\partial\gamma_y},\,\,\,S_z = -\frac{\partial {\hat H}}{\partial \kappa}
\end{align}
 For a chosen set of parameters $(\gamma_x^0, \gamma_y^0, \kappa^0)$,  \eqref{eq:HellmFeynmExp} implies: 
\begin{equation}
\langle S_x^2\rangle(E) = \left[\frac{N}{\Lower{D}(E)}\frac{\partial}{\partial\gamma_x}\Lower U_{(\gamma_x, \gamma_y^0, \kappa^0)}(E)\right]\biggr\rvert_{\gamma_x=\gamma_x^0}
\end{equation}
An analogous expression holds for $S_y^2$ and $S_z$. The calculated spin observables are shown in figure \ref{Plots:SpinObservables}.   The observables all exhibit non-analyticities at energies that correspond to ESQPTS.  Note that these expressions can be generalized to other spin sectors.\\

\section{ Long-Range Interacting 1D Ising Chain\label{sec: Ising}}
In the preceding sections, we demonstrated that 
 the Simon-Lieb inequalities provide a powerful methodology for calculating the density of states of models with all to all interactions. We now explore the utility of this method when applied to models beyond the all-to-all paradigm.
We  specifically consider the  long-range interacting   Ising spin-$\frac 12$ chain  in  a transverse field. This more general family of models has gained a lot of attention through experimental realization via quantum simulation platforms \cite{Britton2012TrappedIonIsing, Browaeys2020ManyBodyRydberg}. Theoretically, this is a technically challenging  and rich problem\cite{HalimehLongRange_2017, HeylLongRange_2018} where known analytical methods break down \cite{Hauke_2013}. The Hamiltonian is 
\begin{equation}\label{eq:lr}
    \hat H = -\sum_{k\neq l}J_{kl}s_{kl}^{zz}-\kappa S_x
\end{equation}
where $s_{kl}^{zz}= s_k^zs_l^z$  and  $S_x=\sum_j s_j^x$ with $s_j^\nu$ denoting a spin-$\frac 12 $ operator. $\kappa$ is the transverse field, and the interaction is described by a power law  
\begin{equation}
    J_{kl} = \frac{\gamma}{|k-l|^\eta}\mathcal{N}^{-1}_{N,\eta}
\end{equation}
Here, the Kac normalization factor for a 1d chain is defined as $\mathcal{N}_{N,\eta}=\sum_{j=1}^N\frac{1}{j^\eta}$ and ensures that energy remains extensive $O(N)$. In this work,  we restrict ourselves to the the truly long range parameter range $\eta\in[0;1)$.\\

The long-range interacting Ising chain breaks two important symmetries of the LMG-Model: the total spin symmetry $[\hat{H}, \Vec{S}^2]\neq 0$ and the permutation symmetry $[\hat{H}, \Pi_{ij}]\neq 0$. Therefore, the Hamiltonian is no longer block diagonal in individual $\text{SU}(2)$ irreps. In the following, we will introduce larger sectors that also encode the permutation structure and block diagonalize the Hamiltonian to utilize our method.

\subsection{Schur-Weyl-Duality and $S_N$ Representations}
The primary idea is to view the joint Hilbert Space $(\mathcal{H}_{\frac{1}{2}}^{\otimes N})$ of the interacting $N$ spin-$\frac{1}{2}$ particles as a representation of both   $\text{SU}(2)$ (rotational) and $S_N$ (permutational) groups. This idea was introduced by Weyl \cite{Weyl1946} and has been used in quantum chemistry \cite{Pauncz1995}, quantum information \cite{Keyl2001}, and more recently to study the Heisenberg chain \cite{Bjornberg2023} . More precisely, we are looking at the joint action $\rho:\text{SU}(2)\times S_N\to \text{U}(\mathcal{H}_{\frac{1}{2}}^{\otimes N})$ defined by: 
\begin{align}
    &\rho(A, \pi)(\ket{\psi_1}\otimes ...\otimes \ket{\psi_n})\nonumber\\&:=T_{\frac{1}{2}}(A)\ket{\psi_{\pi^{-1}(1)}}\otimes...\otimes T_{\frac{1}{2}}(A)\ket{\psi_{\pi^{-1}(n)}}
\end{align}
Where $T_{\frac{1}{2}}$ is the spin-$\frac{1}{2}$ irrep as defined subsection \ref{subsec:SU(2)CoherentStates}. This action rotates the system and permutes the individual components. It has the following useful decomposition into invariant and irreducible subspaces: 
\begin{equation}\label{eq:SchurWeyl}
    \mathcal{H}_{\frac{1}{2}}^{\otimes N}\cong \bigoplus_{s=0}^{\lfloor\frac{N}{2}\rfloor}\mathcal{H}_{\frac{N}{2}-k}\otimes V_{(N-k,k)}
\end{equation}
This is akin to a generalization of the Clebsch-Gordan decomposition: the multiplicity space belonging to each spin-$s$ irrep is endowed with the additional structure of an $S_N$ irrep. The relevant $S_N$ irreps $V_{(N-k,k)}$ are indexed by partitions of $N$ elements into two subsets $\{1,\cdots ,N\}=\{1,\cdots,N-k\}\cup\{1,\cdots,k\}$. This is visualized in the following  Young tableau structure: 
\begin{equation}\label{eq:young}
    V_{(N-k,k)}\cong \ytableausetup{centertableaux}
  \begin{ytableau}
    {} & \none[\cdots] & \none[\cdots] & \none[\cdots]& {} & 
    \none[\quad\scriptstyle\,\Big\}\,N-k] \\
    {} & \none[\cdots] & {} & \none[\scriptstyle\Big\}\,k]
  \end{ytableau}
\end{equation}
We first note that the long range interacting Hamiltonian is not block diagonal in these sectors. In the following, we will study the individual diagonal and off-diagonal blocks. This will allow us to define new sectors where the Hamiltonian will be block diagonal.
\subsection{Highest Spin Sector and Mean Field Analysis}
The starting point of our analysis will be the highest spin sector $s=N/2$. This spin sector has multiplicity one and corresponds to the space $\mathcal{H}_{N/2}\otimes V_{(N,0)}$ in the Schur-Weyl decomposition \eqref{eq:SchurWeyl}. Notably, $V_{(N,0)}$ is the 1d trivial representation of $S_N$, which is why this sector is termed the symmetric sector in the literature. For $\eta=0$ (all-to-all interacting LMG case), this sector becomes an invariant subspace of the Hamiltonian. As we have proven in Appendix \ref{App:Proof LMG GS}, the ground state lies within this sector for ferromagnetic couplings. 
Turning on the interaction range $\eta\in[0;1)$, the ground state will now include contributions from the lower spin sectors. 
As a first step, we  project our Hamiltonian onto the highest spin sector  $\mathcal{D}_1:=P_{\frac{N}{2}}HP_{\frac{N}{2}}$ and then analyze the ensuing corrections introduced by the lower spin sectors.  The principal advantage is that
since $\mathcal{D}_1$ acts on an individual irrep, the Simon-Lieb inequality \eqref{ineq:Simon} holds,  thereby permitting  the use of our method to evaluate the bounds on the density of states. 
The central identity that we  use  for the calculation of the symbols is: 
\begin{align}\label{eq:s_ij identity}
    e^{i\theta S_y}s_{kl}^{zz}e^{-i\theta S_y}&=s_{kl}^{zz}\cos^2\theta+s_{kl}^{xx}\sin^2\theta\nonumber\\&-\frac{\sin 2\theta}{2}(s_{kl}^{zx}+s_{kl}^{xz}) 
\end{align}
This identity can be proven by using the BCH-formula to expand the left hand side. The terms on the right hand side couple different spin sectors. Noting $s^x_k=(s^+_k+s^-_k)/2$, the $\cos^2\theta$ term couples spin difference $\Delta s=0$, the $\sin(2\theta)$ term couples spin difference $\Delta s=1$, and the $\sin^2(\theta)$ term couples spin difference $\Delta s=2$. Consequently (using the definition of the coherent state basis introduced earlier in \eqref{eq:SU(2)CohState}), in the highest spin sector, we have:
\begin{align}
    \bra{\Omega}s_{kl}^{zz}\ket{\Omega}
    =\bra{\Uparrow}e^{i\theta S_y}s_{kl}^{zz}e^{-i\theta S_y}\ket{\Uparrow}=\frac{1}{4}\cos^2\theta
\end{align}
where $\ket{\Uparrow}:=\ket{\uparrow\cdots\uparrow}=\ket{N/2;N/2}$ is the highest weight vector of the spin-$\frac{N}{2}$ irrep. This yields the lower symbol of $\mathcal{D}_1$.  
\begin{align}
    &\Lower{\mathcal{D}_1}(\Omega)=\bra{\Omega}\mathcal{D}_1\ket{\Omega}\nonumber\\
    &=-\frac{\gamma\cos^2\theta}{4\mathcal{N}_{\eta, N}}\sum_{k\neq l}\frac{1}{|k-l|^\eta}-\frac{N\kappa}{2}\sin\theta\cos\varphi
\end{align}
As we are interested in the thermodynamic limit $N\to\infty$, where $\Lower{\mathcal{D}_1}(\Omega)\simeq\Upper{\mathcal{D}_1}(\Omega)$, the above expression tends to: 
\begin{align}
    \mathcal{D}_1(\Omega)&:=\Lower{\mathcal{D}_1}(\Omega)\simeq\Upper{\mathcal{D}_1}(\Omega)\nonumber\\
    &=-\frac{2N\gamma}{2-\eta}\cos^2\theta-\frac{N\kappa}{2}\sin\theta\cos\varphi
\end{align}
This  resembles  an LMG-Type Symbol  where the  parameter $\gamma$ is renormalized by the spatial structure of the coupling. This  predicts the standard Weiss mean field result of a critical  transition when  $2\gamma/(2-\eta)=\kappa$. While this result is exact for the LMG Model, we now address the  corrections stemming from the lower spin sectors. 

\subsection{Correction Through the $s=\frac{N}{2}-1$ Sector}
To obtain the corrections, we now generalize our coherent state basis to explicitly include the permutation symmetry group.
First, note that the $s=\frac{N}{2}-1$ sector is characterized by  the  subspace $\mathcal{H}_{\frac{N}{2}-1}\otimes V_{(N-1,1)}$. The corresponding  $S_N$-irrep (see \eqref{eq:young})has
 dimension $N-1$ and is known as the standard representation in the literature. We again use Perelomov's general construction \cite{Perelomov1986} to construct a coherent state basis for the tensor product $\mathcal{H}_{\frac{N}{2}-1}\otimes V_{(N-1,1)}$. To this end, we pick the highest weight vector with one spin flip: 
\begin{equation}\label{eq:HighestWeightPermutation}
    \ket{\mathbf{1}}:=\ket{\downarrow_1}-\frac{1}{N-1}\sum_{l\neq 1}\ket{\downarrow_l}
\end{equation}
where we have defined $\ket{\downarrow_l}$ as the basis vector with one down spin at position $l$ while all other spins point up. It is straight-forward to verify $S_+\ket{\mathbf{1}}=0$ and $S_z\ket{\mathbf{1}}=(\frac{N}{2}-1)\ket{\mathbf{1}}$. The isotropy subgroup $F$ consists of all rotations around the z-axis (as in the pure $\text{SU}(2)$ case) and additionally all permutations of the last $N-1$ spins: $F=\text{U}(1)\times S_{N-1}$. The coset space now becomes $\text{SU}(2)/\text{U}(1)\times S_N/S_{N-1}$. The permutation group coset space can be intuitively pictured as follows: we pick one spin position $a\in\{1,\cdots ,N\}$, which is where the first spin in equation \eqref{eq:HighestWeightPermutation} is mapped to. The $S_{N-1}$ degree of freedom, which we divide out, permutes the rest of the spins. Thus, the coset space is described by transpositions $\{(1a):a\in \{1,\cdots ,N\} \}$, we can write $\text{SU}(2)/\text{U}(1)\times S_N/S_{N-1}\cong S^2\times \{1,\cdots ,N\}$. Incorporating this coset structure stemming from the permutation symmetry group, we  define a new coherent state basis,  spanning the {}$s=\frac{N}{2}-1$ sector:
\begin{align}
    \ket{\Omega, \mathbf{a}}:= e^{-i\theta S_y}e^{-i\varphi S_z}\Pi_{(1a)}\ket{\mathbf{1}}\\
    = e^{-i\theta S_y}e^{-i\varphi S_z}\ket{\mathbf{a}}
\end{align}
where,
\begin{equation}
    \ket{\mathbf{a}}:=\ket{\downarrow_a}-\frac{1}{N-1}\sum_{l\neq a}\ket{\downarrow_l};\,a\in \{ 1,..., N\}
\end{equation}
These are overcomplete, and their overlap reads:
\begin{equation}\label{eq:PermutiationOverlap}
    \braket{\mathbf{a}|\mathbf{b}}=\frac{N}{N-1}(\delta_{ab}-\frac{1}{N-1})
\end{equation}
The full resolution of  the identity operator on $\mathcal{H}_{\frac{N}{2}-1}\otimes V_{(N-1,1)}$ now takes the form:
\begin{equation}
    \mathds{1}=\frac{(N-1)^2}{N^2}\frac{N-1}{4\pi}\sum_{a=1}^N\int_{S^2}d\Omega\,\ket{\Omega, \mathbf{a}}\bra{\Omega, \mathbf{a}}
\end{equation}
In the $N\to \infty$ limit, 
\begin{equation}
    \mathds{1}\simeq N \sum_{a=1}^N\int_{S^2}\frac{d\Omega}{4\pi}\,\ket{\Omega,\mathbf{a}}\bra{\Omega,\mathbf{a}}
\end{equation}

 To capture the off-diagonal contributions stemming from the lower spin sectors,  we generalize the approach 
 developed in section \ref{subsec:DiagonalSymbols} 
 to now represent operators {\it{between}} different representation spaces. 
For the Hamiltonian of {\eqref{eq:lr}}, we restrict ourselves to operators mapping between two spin representations labeled $s_1$ and $s_2$. A generalization of the results of \cite{Arecchi1972} shows that mappings $O:\mathcal{H}_{s_1}\to \mathcal{H}_{s_2}$ can be represented in the following way:
\begin{equation}\label{eq:off}
    O= C_{s_1,s_2}\int_{S^2} \frac{d\Omega}{4\pi}\,\,\Upper{O}_{s_1,s_2}(\Omega)\ket{\Omega}_{s_2}\bra{\Omega}_{s_1}
\end{equation}
Choosing $C_{s_1,s_2}=\sqrt{(2s_1+1)(2s_2+1)}$, for $s_1=s_2$ we   recover the  standard  definition of an upper symbol.  For $s_1 \ne s_2$, {\eqref{eq:off}}  defines the {\it{off-diagonal upper symbol}}. In what follows, as were are primarily interested in the higher spin sectors, we drop the subscripts in this symbol. In the first off diagonal block $V:=P_{\frac{N}{2}}HP_{\frac{N}{2}-1}$, 
summing over all of the multiplicity space  related by permutations, we obtain the following off-diagonal upper symbol for Hamiltonian operators mapping between spin sectors: 
\begin{equation}
    V\simeq \frac{N}{4\pi}\sum_{a=1}^N\int_{S^2}d\Omega\,\Upsilon(\Omega,a)\ket{\Omega}\bra{\Omega,\mathbf{a}}
\end{equation}
We are interested in the high $N$ asymptotics of $\Upsilon(\Omega, a)$. With the help of equation \eqref{eq:s_ij identity}, we calculate: 
\begin{equation}
    \bra{\Omega}V\ket{\Omega,\mathbf{a}}=\frac{\sin(2\theta)}{4}\Big(\sigma(a)-\frac{1}{N-1}\sum_{m\neq a}\sigma(m)\Big)
\end{equation}
Where we have defined: 
\begin{equation}
\sigma(a)=\sum_{j\neq\mathbf{a}}J_{ja}=\mathcal{N}_{N,\eta}^{-1}\sum_{j\neq a}\frac{\gamma}{|j-a|^{-\eta}}
\end{equation}
This function has the remarkable property that the sum over the permutation degree of freedom vanishes.
\begin{equation}\label{eq:VanishingProperty}
    \sum_{a=1}^N\bra{\Omega}V\ket{\Omega,\mathbf{a}}=0
\end{equation}
Using this, the coherent state overlap of the permutation degree of freedom (equation \eqref{eq:PermutiationOverlap}), and the high $N$ asymptotics of the spin overlap (equation \eqref{eq:SpinOverlapDeltaAsymp}), we obtain:
\begin{equation}
    \Upsilon(\Omega,{a})\simeq\bra{\Omega}V\ket{\Omega,\mathbf{a}}
\end{equation}

Note that when $\eta=0$ as for the LMG, we  recover $V=0$.
We now consider the hamiltonian projected onto the two highest spin sectors $H_{eff}:=(P_{\frac{N}{2}}+P_{\frac{N}{2}-1})H(P_{\frac{N}{2}}+P_{\frac{N}{2}-1})$. We have already obtained $\mathcal{D}_1:=P_{\frac{N}{2}}HP_{\frac{N}{2}}$ and $V:=P_{\frac{N}{2}}HP_{\frac{N}{2}-1}$ in the coherent state bases. Using \eqref{eq:s_ij identity}, to leading order in $N$, the lower symbol corresponding to the missing piece $\mathcal{D}_2:=P_{\frac{N}{2}-1}HP_{\frac{N}{2}-1}$
  does not depend on the permutation degree of freedom. 
 Consequently, 
\begin{equation}
    \mathcal{D}_2\simeq N\int\frac{d\Omega}{4\pi}\sum_{a=1}^N\mathcal{D}_1(\Omega)\ket{\Omega,\mathbf{a}}\bra{\Omega,\mathbf{a}}
\end{equation}
Denoting: 
\begin{align}
     \mathcal{D}(\Omega)&:=\mathcal{D}_1(\Omega) \\
     \mathbf{\Upsilon}(\Omega)&:=\begin{pmatrix} 
         \Upsilon(\Omega, \mathbf{1}) &\cdots & \Upsilon(\Omega, \mathbf{N})
     \end{pmatrix}^T
\end{align}
$H_{eff}$ can be expressed in matrix form as \eqref{eq:EffHamiltonianMatrixForm}. 
\begin{widetext}
    \begin{equation}\label{eq:EffHamiltonianMatrixForm}
    H_{eff}\simeq N\int \frac{d\Omega}{4\pi}\begin{pmatrix}
        \ket{\Omega}&\ket{\Omega,\mathbf{1}}&\cdots&\ket{\Omega, \mathbf{N}}
    \end{pmatrix}  
    \underbrace{\begin{pmatrix}
        \mathcal{D}(\Omega)& \mathbf{\Upsilon}(\Omega)^T \\
        \mathbf{\Upsilon}(\Omega) & \mathcal{D}(\Omega)\mathds{1}_N
    \end{pmatrix}}_{:=A}
    \begin{pmatrix}
        \bra{\Omega}\\\bra{\Omega,\mathbf{1}}\\\vdots\\\bra{\Omega, \mathbf{N}}
    \end{pmatrix}  
    \end{equation}
\end{widetext}
$A$ is the arrowhead matrix, which can be explicitly diagonalized to obtain our desired corrections to the symbol $\mathcal{D}_1(\Omega)$
{} The structure of the hamiltonian above 
only modifies two eigenvalues. Defining,
\begin{align}
    \mathcal{W}(\Omega)&:=\sqrt{\sum_{a=1}^N|\Upsilon(\Omega,a)|^2}\\
    \ket{B(\Omega)}&:=\frac{1}{\mathcal{W}(\Omega)}\sum_{a=1}^N\Upsilon(\Omega,a)\ket{\Omega,\mathbf{a}}\\
    \ket{\Omega, \pm}&:=\frac{1}{\sqrt{2}}\Big[\ket\Omega\pm\ket{B(\Omega)}\Big]
\end{align}
and confining the effective Hamiltonian  to these  modified eigenvectors, we obtain 
\begin{equation}
    H_{eff}=N\sum_{\pm}\int_{S^2}\frac{d\Omega}{4\pi}[\mathcal{D}(\Omega)\pm\mathcal{W}(\Omega)]\ket{\Omega,\pm}\bra{\Omega,\pm} 
\end{equation}
Using equations \eqref{eq:PermutiationOverlap} and \eqref{eq:VanishingProperty}, we see that the modified coherent states satisfy:
\begin{equation}
    \braket{\Omega,\pm|\Omega',\pm}=\frac{1}{2}(\braket{\Omega|\Omega'}_{\frac{N}{2}}+\braket{\Omega|\Omega'}_{\frac{N}{2}-1})
\end{equation}
In the limit of large $N$, the sqaure of these overlaps approaches as  delta function as in equation \eqref{eq:SpinOverlapDeltaAsymp} and the corresponding upper and lower symbols are  asymptotically equal. $\mathcal{W}$ can be recast as 
\begin{equation}
    \mathcal{W}(\Omega)=  \frac{\gamma|\sin2\theta|}{4}C(\eta)
\end{equation}
where 
\begin{equation}
C(\eta)={\sum_{a=1}^N\Big(\sigma(a)-\sum_{a\neq m}\sigma(m)\Big)^2}
\end{equation}
First, we note that  as $\mathcal{D} \propto N$ and $C(\eta)=O(\sqrt{N})$, the  DOS and the ground state energy density remain unaltered in the thermodynamic limit in this subspace.  Finite size corrections do not qualitatively alter the DOS. The same conclusion holds for  the $s=\frac{N}{2}-2$ sector, where the correction is proportional to $\sin^2\theta$; see \eqref{eq:s_ij identity}.  The prediction of ESQPTs within these sectors remains unchanged compared with the LMG. 

The abovementioned scalings of the orders hold for exponents $\eta\in [0;1)$. This class is classified as strongly long range \cite{LongrangeDefenu2023}, where it has been observed both numerically in \cite{Cannas2000} and through more involved analytical methods \cite{Mori2012} that non mean-field effects do not survive the thermodynamical limit. The calculation through our method gives the same result, with considerably less effort. Starting from the sector $s=\frac{N}{2}-2$,  $\eta$  will affect the order of the off-diagonal symbol and, thus, the correction. In the first non-trivial sector $\frac{N}{2}-1$, we only have interaction sums of the type $\sigma(a)$ where the sum runs over  unflipped spins. These sums are renormalized by the Kac-factor, which makes the correction subleading for all $\eta$.  However, the $\frac{N}{2}-2$ sector involves states with two flipped spins whose relative distance will be normalized by the Kac factor.  The ensuing    correction will  then survive the thermodynamic limit for $\eta >1$  and hence a modification of the DOS in the different sectors.   Nonetheless, we see that for $\eta <1$, the models effectively mimic the physics of the LMG in the highest spin sectors, while  for $\eta>1$, the true range of the coupling becomes relevant for the physics. The exact calculations of the corrections and  their impact on  ESQPT's for general $\eta$ go beyond the scope of this paper and will be addressed in future work.

\section{\label{sec:Conc}Conclusion and Outlook}
We have introduced a method to calculate the density of states (DOS) of a system with a general underlying symmetry group within irreducible sectors. This was done by constructing a classical phase space for a general quantum system and identifying two functions on this classical phase space as the lower and upper symbols of the quantum Hamiltonian operator. The inverse Laplace transform of the Simon-Lieb inequality \ref{ineq:Simon} (proved by Simon \cite{Simon1980}) allowed us to calculate the DOS in the high quantum number limit. We further showed that this construction of a classical phase space yields a natural way to calculate microcanonical observables in a formalism akin to classical statistical mechanics. The explicit bounds on the free energy, directly implied by the Simon-Lieb inequality, were shown to yield bounds for the ground-state energy of a general quantum system. 

We have applied all of these general methods to the well-studied LMG model. We first started by identifying the well known ground state QPT in this model. We demonstrated that the phase diagram of the LMG-Model can be decomposed into further sections, investigating the different irregularities of the DOS in various parameter regimes (in the highest spin sector of the thermodynamic limit). In the next step, we applied our method to calculate the microcanonical expectation values to investigate the close relationship between the irregularities of the DOS and the irregularities in the microcanonical expectation values. The calculated DOS and microcanonical expectation values, as well as the phase diagram obtained, have been shown to be in perfect agreement with the analytical results in \cite{Ribeiro_2008}. We further calculated the density of states (DOS) beyond the highest spin sector, obtaining information about the entire spectrum. Here, we were able to benchmark the numerical results in \cite{corps2024} and prove the qualitative similarity of the DOS across different spin-sectors due to the all-to-all nature of the LMG model. In the last section,  generalising the coherent state bases to include both SU(2) and $S_N$ features, we  provided a systematic  framework to obtain  the DOS for   the  ferromagnetic long-range interacting transverse  Ising model. Using this, we showed that only finite size corrections  manifest in the DOS in the highest spin sectors, establishing the robustness of mean-field like physics for interaction exponents less than one.  We leave further exploration of this avenue for future work.  

In principle, though our approach  permits the study of both QPTS and ESQPTs  for any Hamiltonian arising from a compact group $G$, its power, however, lies when applied to models with infinite range interactions. It can be  extended rigorously to study putative ESQPTs in  generalisations of  the SU(2) LMG type Hamiltonians incorporating quartic (and higher order) terms \cite{Magnani_2014, QuantumSpinIce_2014}, terms with mixed interactions \cite{Eckardt_2017}, and more interestingly to the family of SU(3) LMG like models  studied both experimentally and theoretically in \cite{Lin2022main}. 
Lastly, an open question  of great interest is the generalization of the Simon-Lieb inequalities to non-thermal density matrices to study Liouvillian DOS dissipative systems.


\appendix
\newpage
\section{Evaluation of the Volume Integral over $S^2$ in DOS Calculation for LMG}\label{app:IntegralEvaluation}
Our goal is to evaluate the volume integral in equation \eqref{eq:LMG DOS} for the Hamiltonian symbols $f = \Lower H, \Upper H$. 
\begin{align}
    &U_f(E) = Vol_{S^2}\{\Omega\in S^2:f(\Omega)\leq E\}\\ 
    &= \int_{\{f(\Omega)\leq E\} }\frac{d\Omega}{4\pi} = \int_0^{2\pi}d\varphi\int_{\{\theta\in (0;\pi):f(\varphi,\theta)\leq E\}}\sin\theta d\theta\nonumber
\end{align}
The key idea of the calculation is that if we insert $\sin^2\theta = 1-\cos^2\theta$, $\Lower H$, and $\Upper H$ become polynomials of second order in $-\cos\theta$. Defining
\begin{align}
    \Lower A(\varphi) &:= \frac{s(s-1/2)}{N}(\gamma_x\cos^2\varphi+\gamma_y\sin^2\varphi)\\
    \Upper A(\varphi) &:= \frac{(s+1/2)(s+3/2)}{N}(\gamma_x\cos^2\varphi+\gamma_y\sin^2\varphi)\nonumber
\end{align}
and substituting $x = -\cos\theta$, we obtain the polynomials $\Lower P$ and $\Upper P$.
\begin{align}
    \Lower P^\varphi(x) &= x^2\Lower A(\varphi)+xsh-\Lower A(\varphi)-\frac{s}{2N}(\gamma_x+\gamma_y) \\
    \Upper P^\varphi(x) &= x^2\Upper A(\varphi)+x(s+1)h-\Upper A(\varphi)-\frac{s+1}{2N}(\gamma_x+\gamma_y)\nonumber
\end{align}
The volume integral becomes
\begin{equation}
    U_f(E) = \int_0^{2\pi}d\varphi\int_{\{x\in(-1;1):P_f^\varphi(x)\leq E\}}dx
\end{equation}
This can be evaluated by calculating the zeros of the polynomials and making case distinctions.
\section{Proof for the Bounds on the Ground State Energy\label{App:proofGS}}
Here, we prove that inequality \ref{ineq:freeenergies} together with the existence of the limit as $\beta\to\infty$ of the derivatives of free energies implies inequality \ref{ineq:GSbounds}.\\
\textbf{Lemma:} Let $f,g\in C^1(\mathbb{R)}$. Assume $f(x)\geq g(x)$ for all $x$. Moreover, assume that the limits $a:=\lim_{x\to\infty}g'(x)$ and $b:=\lim_{x\to\infty}f'(x)$ exist and are finite. Then $a\geq b$.\\
\textbf{Proof:} Let us suppose $a<b$. Define $\varepsilon$ such that $b-a>\varepsilon$. Take $x_0\in\mathbb{R}$ such that $|g'(x)-a|<\varepsilon/2$ and $|f'(x)-b|<\varepsilon/2$ for all $x\geq x_0$. Moreover, take $\Delta x<\frac{g(x_0)-f(x_0)}{a-b+\varepsilon}$, which is positive. Then:
\begin{align}
    &f(x_0+\Delta x)-g(x_0+\Delta x)=f(x_0)-g(x_0)+\\&\int_{x_0}^{x_0+\Delta x}ds\,f'(s)-g'(s)\\
    &\leq f(x_0)-g(x_0)+\Delta x(a-b+\varepsilon)<0
\end{align}
This is a contradiction, as it implies $f(x_0+\Delta x)>g(x_0+\Delta x)$. Inequality \ref{ineq:GSbounds} follows as a corollary.
\section{Proof: For $\gamma_x,\gamma_y\geq0$, the LMG Ground State lies within the highest spin sector \label{App:Proof LMG GS}}
Here, we will use the ground state energy bounds introduced in subsection \ref{subseq:GSbounds} to prove that the ground state of the LMG-Hamiltonian, within the parameter range $\gamma_x,\gamma_y\geq0$, lies in the highest spin sector. The proof idea is as follows. Since the Hamiltonian is block diagonal with respect to the individual sectors (due to $[\hat H, \vec S^2]=0$ and the permutation symmetry), the ground state must lie in one of the sectors. In other words, the ground state cannot be a mixture of states in different sectors. Let us denote $E_0(s)$ the minimal energy in each sector, i.e. the ground state energy of the Hamiltonian projected onto the spin-$s$ sector. The bounds given in inequality \ref{ineq:GSbounds} hold in every sector. Let us denote the semiclassical bounds in each sector by $\Upper E_0(s):= \min_{\Omega\in S^2}\Upper H_s(\Omega)$ and $\Lower E_0(s):= \min_{\Omega\in S^2}\Lower H_s(\Omega)$. Recall that by inequality \ref{ineq:GSbounds} we have (for all $s$): 
\begin{equation}
    \Upper E_0(s)\leq E_0(s) \leq \Lower E_0(s)
    \label{ineq:GSforalls}
\end{equation}
Our objective will be to show $E_0(s+1)\leq E_0(s)$, which has the immediate consequence that the ground state lies in the highest spin sector. To this end, we will prove $\Lower E_0(s+1)\leq \Upper E_0(s)$, which directly implies $E_0(s+1)\leq E_0(s)$ by inequality \ref{ineq:GSforalls}. We calculate: 
\begin{align}
    \Upper E_0(s)-\Lower E_0(s+1) &= \min_{\Omega\in S^2}\Upper H_s(\Omega)-\min_{\Omega\in S^2}\Lower H_{s+1}(\Omega)
\end{align}
Our objective is to show that this quantity is greater than or equal to zero.
\begin{align}
    &\min_{\Omega\in S^2}\Upper H_s(\Omega)-\min_{\Omega\in S^2}\Lower H_{s+1}(\Omega)\nonumber\\&\geq \min_{\Omega\in S^2}\{\Upper H_s(\Omega)-\Lower H_{s+1}(\Omega)\}
\end{align}
Now we plug in the symbols $H_s(\Omega)$ and $\Lower H_{s+1}(\Omega)$ as calculated in the equation \eqref{eq:hamsymb}.
\begin{align}
    &\Upper H_s(\Omega)-\Lower H_{s+1}(\Omega) = \frac{\gamma_x}{N}\Big[(s+1)-\frac{1}{4}(s+1)(\sin\theta\cos\varphi)^2\Big]\nonumber\\
    &+\frac{\gamma_y}{N}\Big[(s+1)-\frac{1}{4}(s+1)(\sin\theta\sin\varphi)^2\Big]\geq 0
\end{align}
This proves the inequality $\Lower E_0(s+1)\leq \Upper E_0(s)$ and concludes the proof.

\section{Diagonalization of the Berezin Transform for Spin Coherent States}
\label{app: Berezin Transform Diagonalization}
We prove here that spherical harmonics $Y_{lm}$ diagonalize the Berezin Transform with the eigenvalues presented in \eqref{eq:BerezinEigenvalues}. We will view the Berezin Transform as a compact self adjoint operator $\mathcal{B}:L^2(S^2)\to L^2(S^2)$. We are looking for the eigenvectors of this operator. It is easily noticed that the transform commutes with the $\text{SO}(3)$ representation: $(\rho(R)f)(\Omega)=f(R^{-1}\Omega)$. This representation decomposes into irreducible spaces spanned by the spherical harmonics. Since each of these spaces has multiplicity one, the Lemma of Schur implies that these need to be eigenspaces of $\mathcal{B}$. This implies: 
\begin{equation}
    \mathcal{B}Y_{lm}=\lambda_lY_{lm}
\end{equation}
We only need to compute the associated eigenvalues $\lambda_l$. To this end, we choose the spherical function $Y_{l0}(\Omega)=\sqrt{\frac{4\pi}{2l+1}}P_l(\cos\theta)$. Here $P_l$ denotes the Legendre Polynomial. Inserting the definition of the integral transform, one obtains: 
\begin{equation}
    \frac{2s+1}{4\pi}\int_{S^2}d\Omega'\,P_l(\cos\theta')|\braket{\Omega'|\Omega}|^2=\lambda_lP_l(\cos\theta)
\end{equation}
We evaluate both sides at the North Pole $\theta=0$ and use that $P_l(1)=1$. 
\begin{equation}
    \frac{2s+1}{2}\int_{0}^\pi \sin\theta'd\theta'\,P_l(\cos\theta')\Big[\cos\frac{\theta'}{2}\Big]^{4s}=\lambda_l
\end{equation}
We can now use the half angle identity $\cos\frac{\theta'}{2}=\frac{1+\cos\theta'}{2}$, and substitute $x=\cos\theta'$. 
\begin{equation}
    \frac{2s+1}{2}\int_{-1}^1dx\,P_l(x)\Big(\frac{1+x}{2}\Big)^2=\lambda_l
\end{equation}
Substitute $y=\frac{1+x}{2}$ for an even cleaner form.
\begin{equation}
    (2s+1)\int_0^1dy\,P_l(2y-1)y^{2s}=\lambda_l
\end{equation}
Now, we only need to solve this using an explicit representation of Legendre Polynomials. Rodrigues' Formula gives the most direct pathway. In the substituted variable $y$, it reads:
\begin{equation}
    P_l(2y-1)=\frac{(-1)^l}{l!}\frac{d^l}{dy^l}\Big[y^l(1-y)^l\Big]
\end{equation}
We insert this into the integral definition and partially integrate $l$ times to obtain: 
\begin{align}
    \lambda_l=\frac{(2s+1)!(2s)!}{(2s+l+1)!(2s-l)!},\, l\leq 2s
\end{align}
Here, we have used the definition of the Beta Function to reduce the integral into an expression involving factorials. This proves equation \eqref{eq:BerezinEigenvalues}.

\bibliography{apssamp}

\begin{thebibliography}{42}
\bibitem [{Bou()}]{Boundedness}%
  \BibitemOpen
  \href@noop {} {}\bibinfo {note} {The argument for the existence and well-definedness of these values is as follows. The continuity of the representation $T$ implies that the isotropy subgroup $F$ is closed. Thus $\Gamma=G/F$ is compact. Let $h=\Lower H,\Upper H$. We know $h:\Gamma\to\mathbb{R}$ is continuous. Thus the image $h(\Gamma)$ is a compact subset of $\mathbb{R}$}\BibitemShut {NoStop}%
\end{thebibliography}

\end{document}